\documentclass[12pt, twocolumn, numberedappendix, tighten]{aastex62}
\usepackage{natbib}
\usepackage{color} 
\usepackage{morefloats}
\usepackage{gensymb} 

\newcommand{\kms}{\ifmmode {\rm km\ s}^{-1} \else km s$^{-1}$\fi}
\newcommand{\ergs}{\ifmmode {\rm erg\ s}^{-1} \else erg s$^{-1}$\fi}
\newcommand{\ergscm}{\ifmmode {\rm erg\ s}^{-1} \else erg s$^{-1}$ cm$^{-2}$\fi}
\newcommand{\Msun}{\ifmmode {\rm M}_{\odot} \else $M_{\odot}$\fi }
\newcommand{\Lsun}{\ifmmode {\rm L}_{\odot} \else L$_{\odot}$\fi}
\newcommand{\qo}{\ifmmode q_{\rm o} \else $q_{\rm o}$\fi}
\newcommand{\Ho}{\ifmmode H_{\rm o} \else $H_{\rm o}$\fi}
\newcommand{\ho}{\ifmmode h_{\rm o} \else $h_{\rm o}$\fi}

\newcommand{\vFWHM}{\ifmmode v_{\mbox{\tiny FWHM}} \else
                    $v_{\mbox{\tiny FWHM}}$\fi}
\newcommand{\CCF}{\ifmmode F_{\it CCF} \else $F_{\it CCF}$\fi}
\newcommand{\ACF}{\ifmmode F_{\it ACF} \else $F_{\it ACF}$\fi}
\newcommand{\Halpha}{\ifmmode {\rm H}\alpha \else H$\alpha$\fi}
\newcommand{\Hbeta}{\ifmmode {\rm H}\beta \else H$\beta$\fi}
\newcommand{\Hgamma}{\ifmmode {\rm H}\gamma \else H$\gamma$\fi}
\newcommand{\Hdelta}{\ifmmode {\rm H}\delta \else H$\delta$\fi}
\newcommand{\Lya}{\ifmmode {\rm Ly}\alpha \else Ly$\alpha$\fi}
\newcommand{\Lyb}{\ifmmode {\rm Ly}\beta \else Ly$\beta$\fi}
\newcommand{\HeI}{\ifmmode {\rm He}\,{\sc i}\,\lambda5876 \else 
	          He\,{\sc i}\,$\lambda5876$\fi}
\newcommand{\HeII}{\ifmmode {\rm He}\,{\sc ii}\,\lambda4686 \else 
	           He\,{\sc ii}\,$\lambda4686$\fi}
\newcommand{\heii}{He\,{\sc ii}}
\newcommand{\mgii}{Mg\,{\sc ii}}

\newcommand{\ciii}{\ifmmode {\rm C}\,{\sc iii} \else C\,{\sc iii}\fi}
\newcommand{\civ}{C\,{\sc iv}}

\newcommand{\mbh}{$M_{\rm BH}$\ }

\newcommand{\rblr}{$R_{\rm BLR}$\ }

\newcommand{\radlum}{$R_{\rm BLR} - L$}
\newcommand{\javelin}{{\tt JAVELIN}\ }
\newcommand{\cream}{{\tt CREAM}\ }

\newcommand{\numlags}{52\ }
\newcommand{\numgold}{18\ }
\newcommand{\numneg}{5\ } 

\shortauthors{Grier et al.}

\begin{document}

\title{The Sloan Digital Sky Survey Reverberation Mapping Project: Initial \civ \  Lag Results from Four Years of Data}

\author{C.~J.~Grier}
\affiliation{Department of Astronomy and Astrophysics, Eberly College of Science, The Pennsylvania State University, 525 Davey Laboratory, University Park, PA 16802}
\affiliation{Institute for Gravitation \& the Cosmos, The Pennsylvania State University, University Park, PA 16802}
\affiliation{Steward Observatory, The University of Arizona, 933 North Cherry Avenue, Tucson, AZ 85721, USA} 

\author{Yue~Shen}
\altaffiliation{Alfred P. Sloan Research Fellow}
\affiliation{Department of Astronomy, University of Illinois at Urbana-Champaign, Urbana, IL 61801, USA} 
\affiliation{National Center for Supercomputing Applications, University of Illinois at Urbana-Champaign, Urbana, IL 61801, USA} 

\author{Keith~Horne}
\affiliation{SUPA Physics and Astronomy, University of St. Andrews, Fife, KY16 9SS, Scotland, UK} 

\author{W.~N.~Brandt}
\affiliation{Department of Astronomy and Astrophysics, Eberly College of Science, The Pennsylvania State University, 525 Davey Laboratory, University Park, PA 16802}
\affiliation{Institute for Gravitation \& the Cosmos, The Pennsylvania State University, University Park, PA 16802}
\affiliation{Department of Physics, The Pennsylvania State University, University Park, PA 16802, USA}

\author{J.~R.~Trump}
\affiliation{Department of Physics, University of Connecticut, 2152 Hillside Road, Unit 3046, Storrs, CT 06269, USA}

\author{P.~B.~Hall}
\affiliation{Department of Physics and Astronomy, York University, Toronto, ON M3J 1P3, Canada}

\author{K.~Kinemuchi} 
\affiliation{Apache Point Observatory and New Mexico State University, P.O. Box 59, Sunspot, NM, 88349-0059, USA}

\author{David~Starkey} 
\affiliation{SUPA Physics and Astronomy, University of St. Andrews, Fife, KY16 9SS, Scotland, UK} 
\affiliation{Department of Astronomy, University of Illinois at Urbana-Champaign, Urbana, IL 61801, USA}

\author{D.~P.~Schneider}
\affiliation{Department of Astronomy and Astrophysics, Eberly College of Science, The Pennsylvania State University, 525 Davey Laboratory, University Park, PA 16802}
\affiliation{Institute for Gravitation \& the Cosmos, The Pennsylvania State University, University Park, PA 16802}

\author{Luis~C.~Ho}
\affiliation{Kavli Institute for Astronomy and Astrophysics, Peking University, Beijing 100871, China} 
\affiliation{Department of Astronomy, School of Physics, Peking University, Beijing 100871, China} 

\author{Y.~Homayouni}
\affiliation{Department of Physics, University of Connecticut, 2152 Hillside Rd Unit 3046, Storrs, CT 06269, USA}

\author{Jennifer~I-Hsiu~Li}
\affiliation{Department of Astronomy, University of Illinois at Urbana-Champaign, Urbana, IL 61801, USA}

\author{Ian~D.~McGreer}
\affiliation{Steward Observatory, The University of Arizona, 933 North Cherry Avenue, Tucson, AZ 85721, USA}

\author{B.~M.~Peterson}
\affiliation{Department of Astronomy, The Ohio State University, 140 W 18th Avenue, Columbus, OH 43210, USA}
\affiliation{Center for Cosmology and AstroParticle Physics, The Ohio State University, 191 West Woodruff Avenue, Columbus, OH 43210, USA}
\affiliation{Space Telescope Science Institute, 3700 San Martin Drive, Baltimore, MD 21218, USA }


\author{Dmitry Bizyaev}
\affiliation{Apache Point Observatory and New Mexico State University, P.O. Box 59, Sunspot, NM, 88349-0059, USA}
\affiliation{Sternberg Astronomical Institute, Moscow State University, Moscow, Russia}

\author{Yuguang Chen}
\affiliation{California Institute of Technology, 1200 E California Blvd., MC 249-17, Pasadena, CA 91125, USA}  

\author{K.~S.~Dawson}
\affiliation{Department of Physics and Astronomy, University of Utah, 115 S. 1400 E., Salt Lake City, UT 84112, USA} 

\author{Sarah~Eftekharzadeh}
\affiliation{Department of Physics and Astronomy, University of Utah, 115 S. 1400 E., Salt Lake City, UT 84112, USA}

\author{Yucheng Guo}
\affiliation{Department of Astronomy, School of Physics, Peking University, Beijing 100871, China} 

\author{Siyao~Jia}
\affiliation{Department of Astronomy, University of California, Berkeley, CA 94720, USA}

\author{Linhua~Jiang}
\affiliation{Kavli Institute for Astronomy and Astrophysics, Peking University, Beijing 100871, China}

\author{Jean-Paul Kneib}
\affiliation{Institute of Physics, Laboratory of Astrophysics, Ecole Polytechnique F\'ed\'erale de Lausanne (EPFL), Observatoire de Sauverny, 1290 Versoix, Switzerland}
\affiliation{Aix Marseille Universit\'e, CNRS, LAM (Laboratoire d'Astrophysique de Marseille) UMR 7326, 13388, Marseille, France}

\author{Feng Li} 
\affiliation{School of Mathematics and Physics, Changzhou University, Changzhou 213164, China} 

\author{Zefeng Li} 
\affiliation{Department of Astronomy, School of Physics, Peking University, Beijing 100871, China} 

\author{Jundan Nie}
\affiliation{Key Laboratory of Optical Astronomy, National Astronomical Observatories, Chinese Academy of Sciences, Beijing 100012, China}

\author{Audrey Oravetz}
\affiliation{Apache Point Observatory and New Mexico State University, P.O. Box 59, Sunspot, NM, 88349-0059, USA}

\author{Daniel Oravetz}
\affiliation{Apache Point Observatory and New Mexico State University, P.O. Box 59, Sunspot, NM, 88349-0059, USA}

\author{Kaike Pan}
\affiliation{Apache Point Observatory and New Mexico State University, P.O. Box 59, Sunspot, NM, 88349-0059, USA}

\author{Patrick Petitjean} 
\affiliation{Institut d'Astrophysique de Paris, Sorbonne Universit\'e and CNRS,
   98bis Boulevard Arago, 75014, Paris, France}

\author{Kara~A.~Ponder}
\affiliation{Berkeley Center for Cosmological Physics,
    University of California Berkeley,
    341 Campbell Hall, Berkeley, CA 94720, USA  }

\author{Jesse~Rogerson}
\affiliation{Canada Aviation and Space Museum, 11 Aviation Parkway, Ottawa, ON, K1K 4Y5, Canada}
\affiliation{Department of Physics and Astronomy, York University, Toronto, ON M3J 1P3, Canada}

\author{M.~Vivek} 
\affiliation{Department of Astronomy and Astrophysics, Eberly College of Science, The Pennsylvania State University, 525 Davey Laboratory, University Park, PA 16802}
\affiliation{Institute for Gravitation \& the Cosmos, The Pennsylvania State University, University Park, PA 16802}

\author{Tianmen Zhang} 
\affiliation{Key Laboratory of Optical Astronomy, National Astronomical Observatories, Chinese Academy of Sciences, Beijing 100012, China}
\affiliation{School of Astronomy and Space Science, University of Chinese Academy of Sciences}

\author{Hu Zou} 
\affiliation{Key Laboratory of Optical Astronomy, National Astronomical Observatories, Chinese Academy of Sciences, Beijing 100012, China}
 

\begin{abstract}
We present reverberation-mapping lags and black-hole mass measurements using the \civ $\lambda1549$ broad emission line from a sample of 349 quasars monitored as a part of the Sloan Digital Sky Survey Reverberation Mapping Project. Our data span four years of spectroscopic and photometric monitoring for a total baseline of 1300 days. We report significant time delays between the continuum and the \civ$\lambda$1549 emission line in \numlags quasars, with an estimated false-positive detection rate of 10\%. Our analysis of marginal lag measurements indicates that there are on the order of $\sim$100 additional lags that should be recoverable by adding more years of data from the program.  
We use our measurements to calculate black-hole masses and fit an updated \civ \ radius-luminosity relationship. Our results significantly increase the sample of quasars with \civ \ RM results, with the quasars spanning two orders of magnitude in luminosity toward the high-luminosity end of the \civ \ radius--luminosity relation. In addition, these quasars are located at among the highest redshifts ($z \approx$ 1.4--2.8) of quasars with black hole masses measured with reverberation mapping. 
This work constitutes the first large sample of \civ \ reverberation-mapping measurements in more than a dozen quasars, demonstrating the utility of multi-object reverberation mapping campaigns.
\end{abstract}

\keywords{galaxies: active --- galaxies: nuclei --- quasars: general --- quasars: emission lines}

\section{INTRODUCTION}
\label{introduction}
Supermassive black holes (SMBHs) are nearly ubiquitous in massive galaxies across the Universe, and their masses have been shown to be correlated with a variety of properties of the galaxies in which they reside (e.g.,  \citealt{Kormendy95}; \citealt{Magorrian98}; \citealt{Ferrarese00}; \citealt{Gebhardt00a}; \citealt{Gultekin09}). As a consequence, theories and simulations regarding the evolution of galaxies must include SMBHs; explaining how SMBHs grew to their observed masses and how they are connected to their host galaxies is a critical component of galaxy evolution models. Accurate measurements of SMBH masses are therefore of paramount importance to successfully explain the connection between galaxies and their SMBHs across the observable Universe. 

In nearby galaxies, black-hole mass ($M_{\rm BH}$) measurements can be obtained from observations of stellar and gas dynamics near the center of the galaxy (e.g., \citealt{McConnell13}). However, this approach is currently infeasible for distant galaxies; to determine \mbh in galaxies beyond the local universe, we use active galactic nuclei (AGN). Assuming that the broad emission lines observed in Type~1 AGN are emitted by gas whose motion is dominated by the gravitational potential of the central SMBH, one can use this gas to obtain \mbh measurements. However, as the broad line-emitting regions (BLR) in most AGN are too small to directly resolve with current technology (see \citealt{Gravity18} for the only exception thus far), there are limited opportunities to learn about the size and structure of the BLR. Reverberation mapping (RM) is the primary technique employed for this (the other being gravitational microlensing; e.g., \citealt{Morgan10}; \citealt{Mosquera13}). 

RM uses the variability of AGN to obtain BLR information: Variations in the continuum flux (generally assumed to be emitted close to the SMBH) are echoed by gas in the BLR, with the signal from the BLR delayed by the light-travel time between the continuum-emitting source and the BLR gas (e.g., \citealt{Blandford82}; \citealt{Peterson04}).  Measuring this time delay determines the distance between these two regions, which yields a characteristic radius for the BLR, $R_{\rm BLR}$. This measurement can be combined with a characterization of the virial velocity of the gas, $\Delta V$, which is assumed to be related to the width of the emission line, to yield a black hole mass: 
\begin{equation} 
M_{\rm BH} = \frac{fR_{\rm BLR}\Delta V^2}{G}, 
\label{eq:mbh}
\end{equation} 
where $f$ is a dimensionless factor that accounts for the geometry, orientation, and kinematics of the BLR. 
 
In theory, RM measurements can be made using any suitably strong broad emission lines arising from gas that reverberates in response to the continuum and is in virial motion around the SMBH. Thus far, most ground-based efforts have been focused on the \Hbeta \ emission line, which falls in the optical range in local AGN, and additional strong optical lines such as \Halpha, \Hgamma, and \heii$\lambda4686$. Attention has also been given to the \civ $\lambda$1549 and \mgii$\lambda$2798 emission lines, which are often quite strong, and lie within the optical range of many ground-based spectrographs for higher-redshift quasars. To date, on the order of ~100 AGN have RM measurements (e.g., \citealt{Kaspi00, Kaspi05, Peterson04, Bentz09c, Bentz10a, Denney10, Grier12b, Du14, Du16a, Du16b, Barth15, Hu15, Grier17b, Lira18}). 

RM measurements of local AGN have established a tight correlation between \rblr and the luminosity of the AGN (e.g., \citealt{Kaspi00}; \citealt{Kaspi05}; \citealt{Bentz13}), with $R \propto \sqrt{L}$, consistent with basic photoionization expectations. This relation allows the estimation of \rblr from a single spectrum, enabling \mbh estimates (hereafter referred to as single-epoch, or SE, masses) for a large number of quasars for which RM campaigns are impractical (e.g., \citealt{Shen11}). The current \Hbeta\ \radlum \ relationship is fairly well calibrated (\citealt{Bentz13}), although there is a dearth of measurements at the high-luminosity end of the relation. The sample included in the most recent calibration of this relation is composed of $\sim$40 nearby ($z < 0.3$), low-luminosity AGN that may not be representative of the general AGN/quasar population. Recent studies by \cite{Du16a} and \cite{Grier17b} find many objects below the measured relation, although the origin of this phenomenon is still currently under investigation and selection effects are likely relevant in some cases (e.g., \citealt{Li19}; Fonseca et al. in preparation). 

Many studies have focused on the \civ$\lambda$1549 emission line because it is one of the few strong lines in the ultraviolet (UV), making \mbh measurements in higher-redshift quasars feasible from the ground. The status of the \civ\ emission line with regards to measuring \mbh is complex: \civ\ frequently exhibits a blueshifted component reminiscent of outflows, and has been found to have significant non-reverberating components (e.g., \citealt{Gaskell82}; \citealt{Korista95}; \citealt{Richards11}; \citealt{Denney12}), though it has been suggested that many of the reported blueshifts are affected by incorrect redshift measurements (\citealt{Denney16b}). In addition, these properties depend on luminosity  --- i.e., the blueshift is observed primarily in higher-luminosity quasars, and recent velocity-resolved RM results of the local Seyfert galaxy NGC\,5548 (\citealt{Derosa15}; Horne et al., in preparation) show signatures indicative of a Keplerian disk with gas in virial motion rather than evidence for outflowing gas. Possibly as a consequence of the above issue, differences have been reported between the FWHM of \civ\ and the FWHM of \Hbeta\ (\citealt{Baskin05}; \citealt{Shang07}; \citealt{Netzer07}; \citealt{Trakhtenbrot12}; \citealt{Shen12}; \citealt{Shen13}), with \civ\ sometimes showing narrower widths than \Hbeta. This has been interpreted as possible evidence against a simple radially stratified BLR that RM studies generally support (e.g., \citealt{Peterson93}; \citealt{Korista95}). These issues have raised concerns over the suitability of \civ\ for SE \mbh estimates, though some studies suggest that data quality is the major issue rather than \civ\ itself (e.g., \citealt{Vestergaard06}; \citealt{Denney12}). Several corrections have been proposed to address these various issues and allow \civ \ to continue be used as a SE estimator (e.g., \citealt{Assef11}; \citealt{Denney12}; \citealt{Runnoe13}; \citealt{Brotherton15}; \citealt{Coatman17}).  With or without these corrections, \civ\ has continued to be used to estimate \mbh in large numbers of sources (e.g., \citealt{Shen11}). 

Despite all of these potential issues, \civ\ can still be used for RM \mbh measurements, as RM methods make use of the root-mean-square (RMS) line profile, which includes only the part of the \civ\ line that does reverberate. However, RM measurements of the \civ \ emission line are difficult to obtain: First, measurements in local galaxies require the use of space telescopes, as rest-frame \civ \ lies in the UV and is not accessible from the ground. Secondly, in higher-redshift, more luminous quasars, the expected observed lags are on the order of years (due to cosmological time dilation), making them impossible to measure in a single observing season and requiring long-term, logistically difficult observing campaigns. It is for these reasons that \civ \ RM measurements are far more scarce than \Hbeta \ RM measurements. Thus far there have been only $\sim$15--18 \civ \ robust RM lag measurements that are used to calibrate the \civ\ \radlum\ relation (\citealt{Peterson04} and references therein; \citealt{Peterson05}; \citealt{Kaspi07}; \citealt{Trevese14}; \citealt{Derosa15}; \citealt{Lira18}; \citealt{Hoormann19}), though there were some earlier reports of \civ\ lag detections of varying quality (e.g., \citealt{Gaskell86}; \citealt{Clavel89}; \citealt{Koratkar89, Koratkar91a}). The most recently measured \radlum \ relations for the \civ \ emission line (\citealt{Lira18}; \citealt{Hoormann19}) still contain relatively few measurements compared to the \Hbeta \ relation, and there are large ranges of luminosities along that relation for which there are no published measurements. 



We have embarked on a large-scale, multi-object RM campaign called the Sloan Digital Sky Survey Reverberation Mapping Project (SDSS-RM; \citealt{Shen15a}), one of the major goals of which is to measure \civ \ lags in a large sample of quasars over a range of luminosities and redshifts. 
SDSS-RM began in 2014 as an ancillary program within the SDSS-III Baryon Oscillation Spectroscopic Survey (BOSS; \citealt{Eisenstein11}; \citealt{Dawson13}) and has continued to acquire spectra thereafter as a part of the SDSS-IV eBOSS program (\citealt{Dawson16}; \citealt{Blanton17}). Spectra of 849 quasars are obtained each observing season between January and July with the SDSS 2.5 m telescope (\citealt{Gunn06}), and accompanying photometric data are acquired with the 3.6 m Canada-France-Hawaii Telescope (CFHT) and the Steward Observatory 2.3 m Bok telescope. Observations will continue to be taken through 2020. The main goals of the program are to obtain RM measurements using the \Hbeta, \mgii, and \civ \ emission lines for quasars over a wide range of redshifts; however, a wide variety of science topics can be (and have been) addressed with the rich dataset provided by the SDSS-RM program, ranging from studies of quasar host galaxies, to broad absorption-line variability, to emission-line properties, to general quasar variability (e.g., \citealt{Shen15b}; \citealt{Grier15}; \citealt{Sun15}; \citealt{Matsuoka15}; \citealt{Denney16a}; \citealt{Shen16}; \citealt{Yue18}; \citealt{Hemler19}; \citealt{Homayouni19}). 

We here present \civ \ RM results from the SDSS-RM quasar sample using data taken during the first four years of the program (2014--2017). We present our quasar sample and the data used in our study in Section~\ref{sec:data}. In Section~\ref{sec:timeseries}, we describe the methodology used for the various measurements, and in Section~\ref{sec:discussion} we discuss our results and their implications. We conclude in Section~\ref{sec:summary} with a summary of our main results. Throughout this article, we adopt a $\Lambda$CDM cosmology with $\Omega_{\Lambda}$~=~0.7, $\Omega_{M}$~=~0.3, and $h$~=~0.7. 

\section{DATA AND DATA PROCESSING} 
\label{sec:data}

\subsection{The Quasar Sample} 
\label{sec:sample}
The parent sample of quasars consists of the 849 quasars monitored in the SDSS-RM field (details on this sample are provided by \citealt{Shen18}). We first restrict our sample to the 492 quasars with $z > 1.3$; i.e., quasars with observed-frame wavelength coverage of the \civ\ emission line in the BOSS spectra. 

\begin{figure}
\begin{center} 
\includegraphics[scale = 0.45, trim = 5 0 0 0, clip]{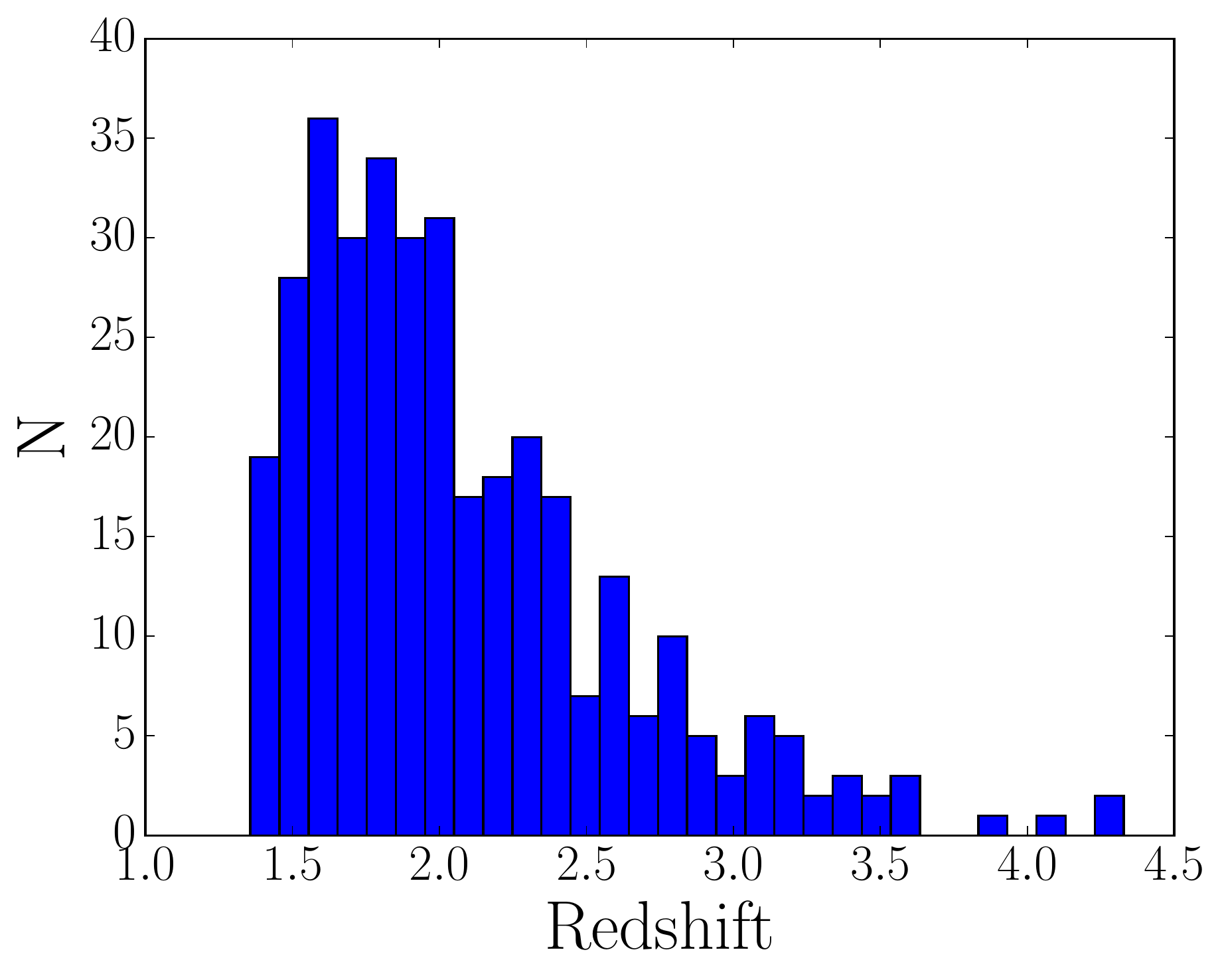}
\includegraphics[scale = 0.45, trim = 5 0 0 0, clip]{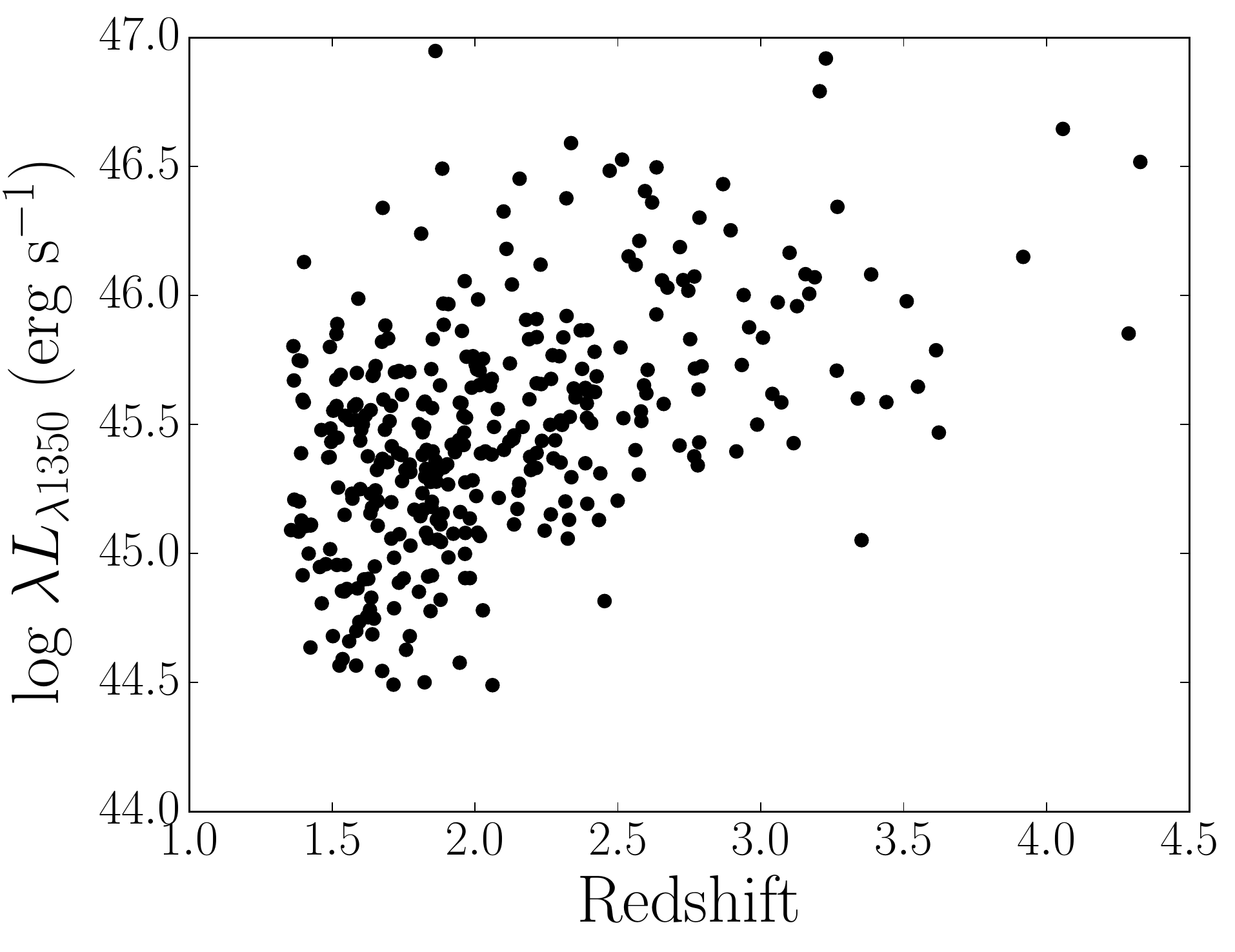}
\includegraphics[scale = 0.45, trim = 5 0 0 0, clip]{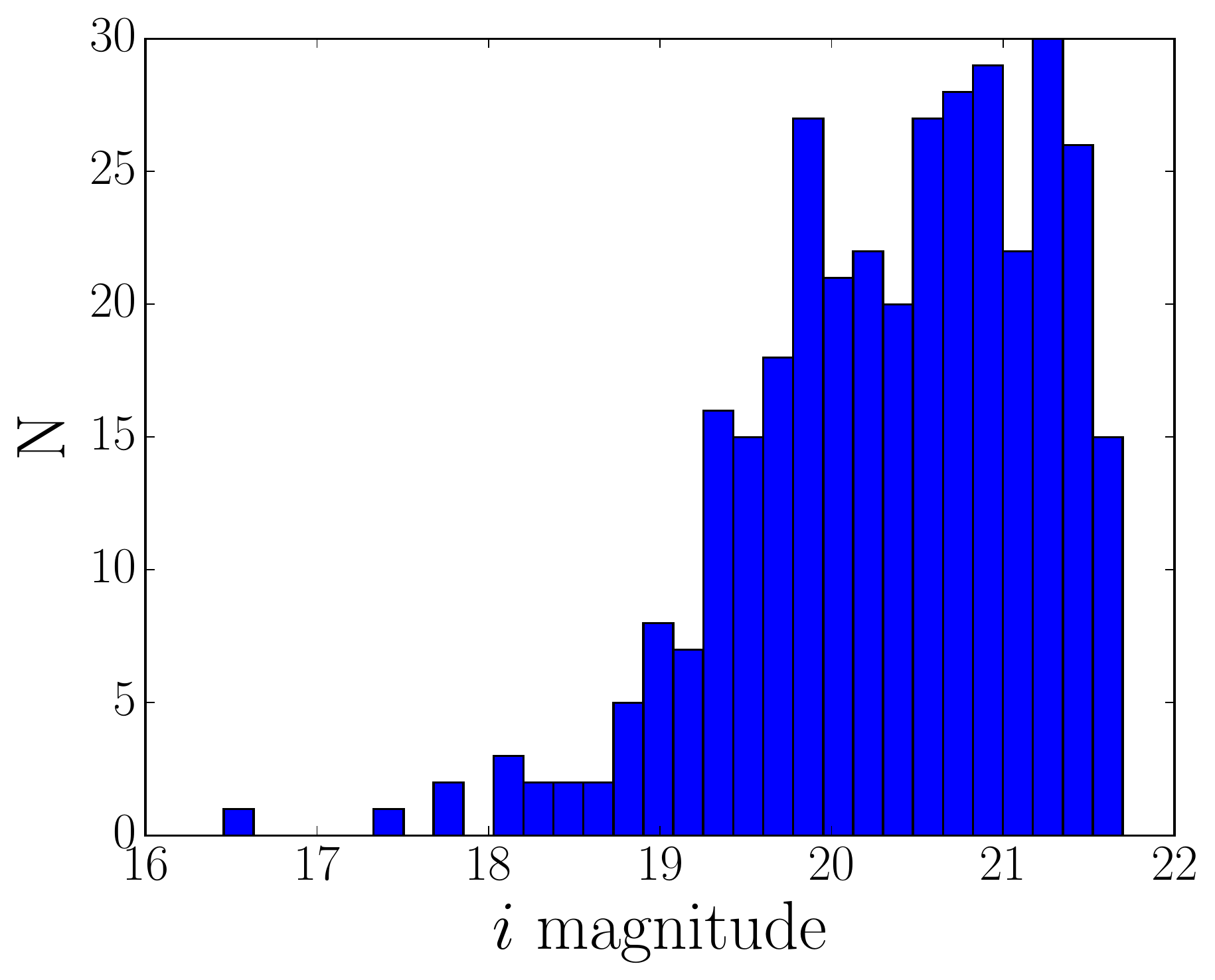}
\caption{The distributions of various properties of our quasar sample. From top to bottom: The redshift distribution,  $\lambda$log$L_{\lambda1350}$ (the continuum luminosity at 1350\,\AA) vs. redshift, and the distribution of $i$-magnitude. All quantities were measured by \cite{Shen18}.} 
\label{fig:sample} 
\end{center} 
\end{figure}

In many sources, however, the \civ \ emission line was not sufficiently variable to obtain RM measurements. Before performing our analysis, we thus first excluded sources whose \civ \ emission lines did not show significant variability over the span of our observations. To characterize the variability, we measured the \civ \ light curve variability signal-to-noise ratio using the quantity SNR2, which is an output from the PrepSpec software (see Section~\ref{sec:prepspec} for a discussion of PrepSpec). SNR2 is defined as $\sqrt{\chi^2 - {\rm DOF}}$, where $\chi^2$ is calculated against the average of the light curve flux (using the measurement uncertainties of the light curves $\sigma_{i}$), and DOF is the degrees of freedom, which is equal to the number of points in the light curve  $-~1$. 
Larger values of SNR2 indicate that the null-hypothesis model of no variability is a poor description of the emission-line light curve, while smaller values indicate that the light curve is consistent with zero variability. We require that SNR2 of the \civ\ emission line is greater than 20 for a quasar to be included in our sample (this number was chosen based on visual inspection of the PrepSpec fits, light curves, and RMS residual line profiles). This criterion produced a final sample of 349 quasars, with redshifts ranging from 1.35 to 4.32. Basic information on these quasars is provided in Table~1, and Figure~\ref{fig:sample} displays the distributions of redshift, $i$-mag, and luminosity of the quasars in our final sample. 

\subsection{Spectroscopic Data}
\label{sec:prepspec} 
\begin{figure}
\begin{center} 
\includegraphics[scale = 0.335, trim = 5 0 0 0, clip]{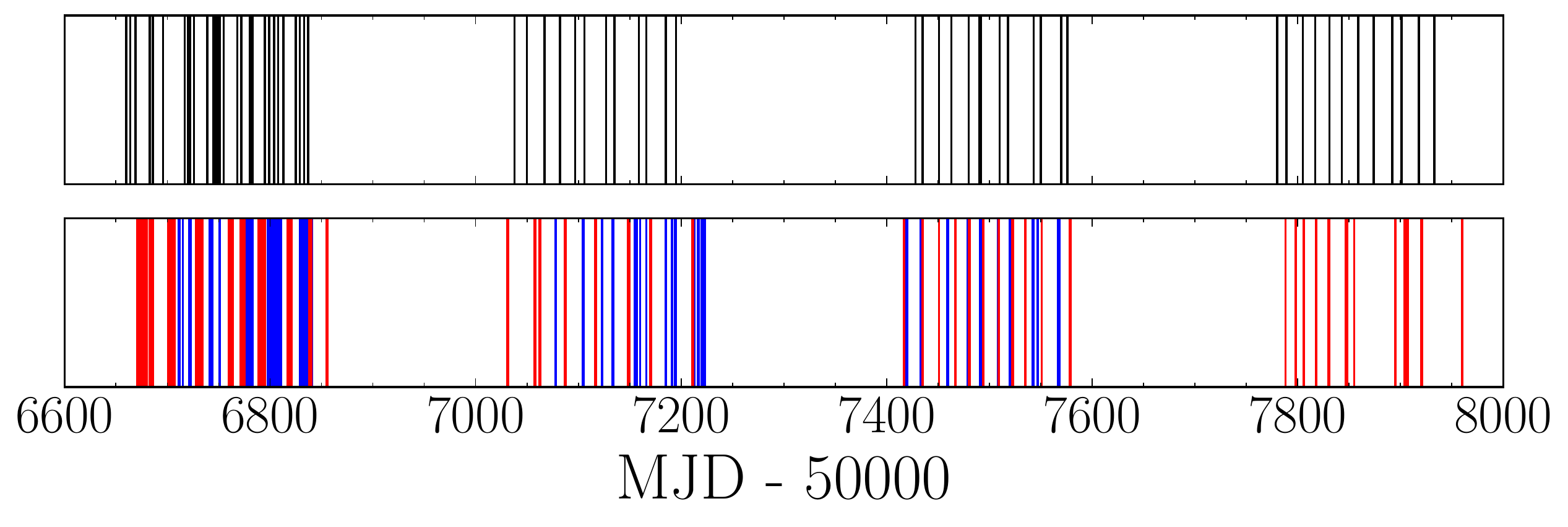}
\caption{The distribution of MJD for the 2014--2017 spectroscopic observations from SDSS (top panel) and photometric observations from the Bok and CFHT (bottom panel).  Each vertical line represents an observed epoch. Black lines indicate SDSS spectroscopic observations, blue lines represent CFHT observations, and red lines indicate Bok observations. The large spacings between sets of lines highlight the seasonal gap between each observing year.} 
\label{fig:cadence} 
\end{center} 
\end{figure}

We obtained the spectra used in this study during the first four years of observations for the SDSS-RM campaign (e.g., \citealt{Shen15a}), which monitors 849 quasars with $i <$ 21.7 at redshifts ranging from 0.1 to 4.5. The spectra were acquired with the BOSS spectrograph (\citealt{Dawson13}; \citealt{Smee13}), which covers a wavelength range of $\sim$3560--10400\,\AA. The spectrograph has a spectral resolution of R$\sim$2000 and the data are binned to 69 km~s$^{-1}$ per pixel. We obtained a total of 68 epochs between 2014 January and 2017 July, with observations taken between January--July in each year only, leaving a 6-month gap between observing seasons. The first year of SDSS-RM monitoring yielded 32 spectroscopic epochs and the additional three years of monitoring yielded 12 epochs each. Figure~\ref{fig:cadence} displays the observing cadence for the observations.  

The 2014 spectra were processed using the standard SDSS-III pipeline (version 5\_7\_1); data from the subsequent years were processed using the updated SDSS-IV eBOSS reduction pipeline (version 5\_10\_1). We then further processed all spectra using a custom flux-calibration scheme described by \cite{Shen15a}, which improves the spectrophotometric calibrations by using additional standard stars observed on the plate. 

To enhance further the relative flux calibration of the data, we employed a custom procedure using proprietary software referred to as PrepSpec (this code is described in detail by \citealt{Shen15a}; \citealt{Shen16}; Horne et al.\ in preparation). PrepSpec models the spectra using a variety of different components and applies a time-dependent flux correction that is calculated by using the narrow emission lines (when present) as a calibrator. The correction assumes that there is no intrinsic variability in the fluxes of the narrow emission lines over the course of the campaign --- some observations of long-term changes in narrow-line flux in local, low-luminosity sources have been reported (e.g., NGC\,5548; \citealt{Peterson13}), but simple luminosity scaling from NGC\,5548 predicts narrow-line variability timescales of $>30$ rest-frame years in our quasars. 

The PrepSpec model includes intrinsic variations in the continuum and broad emission lines, and the model is optimized to fit all of the spectra of an object simultaneously. In addition to the intrinsic variability of the continuum and emission lines, PrepSpec also accounts for variations in seeing and small shifts in the wavelength solution. Various spectral measurements from PrepSpec using the first year of data only are presented by \cite{Shen18}. 

We use PrepSpec to improve our flux calibrations and subsequently to produce measurements of line fluxes, line widths, mean/RMS profiles, and light curves for each emission line (and various continuum regions, depending on the wavelength ranges accessible for each object). We convolve our PrepSpec-corrected spectra with the SDSS filter response curves (\citealt{Fukugita96}; \citealt{Doi10}) to produce $g$- and $i$-band synthetic photometry for each quasar. To estimate the uncertainties in the synthetic photometric fluxes, we sum in quadrature the spectral uncertainties and the errors in the flux-correction factors reported by PrepSpec. 

Before further analysis, we first removed any suspect epochs and outliers from our spectroscopic light curves. The seventh epoch is a significant outlier in a large fraction of the light curves;  following \cite{Grier17b}, we remove this epoch from all of our spectroscopic light curves. In addition, there are occasional spectra (roughly 4\% of epochs) that have zero flux or are significant low-flux outliers in the light curves (these are cases where the BOSS spectrograph fibers were not plugged in correctly or the SDSS pipeline failed to extract a proper spectrum). We excluded all points with zero flux and those that were offset from the median flux by more than five times the normalized median absolute deviation of the light curve (NMAD; \citealt{Maronna06}). 

\subsection{Photometric Data}    
To improve the cadence of our continuum light curves, we also monitored the SDSS-RM field in the $g$ and $i$ bands with the Steward Observatory Bok 2.3m telescope on Kitt Peak from 2014--2017 and the 3.6m CFHT on Maunakea from 2014--2016. We used the Bok/90Prime instrument (\citealt{Williams04}) for our observations, which has a 1\degree $\times$ 1\degree\ field of view, mapping the observations onto a 4k $\times$ 4k CCD with a plate scale of 0.45\arcsec pixel$^{-1}$. 
On the CFHT, we used the MegaCam instrument (\citealt{Aune03}), which has a similar 1\degree $\times$ 1\degree\ field of view and a pixel scale of 0.187\arcsec. 
The observing cadence of the photometric observations is provided in Figure~\ref{fig:cadence}. 

Following \cite{Grier17b}, we adopted the image subtraction method as implemented in the software package ISIS (\citealt{Alard98}; \citealt{Alard00}) to produce the photometric light curves. The basic steps are as follows: (1) The images are aligned; (2) The images with the best seeing, transparency, and sky background are used to create a reference image; (3) For each epoch, the reference image point-spread function (PSF) is altered to match that of the epoch, and a flux-calibration scale factor is applied to the target image; (4) The epoch and the reference image are subtracted, yielding a ``difference" image that has the same flux calibration as the reference image; (5) A residual-flux light curve is produced by placing a PSF-weighted aperture over each source to measure the flux in the subtracted image.  

We performed the image subtraction separately for each individual telescope, field, filter, and CCD to obtain $g$- and $i$-band light curves for each quasar. Before further analysis, we removed problematic epochs from the light curves, such as epochs where the source fell on or near the edge of the detector, epochs where the sources were saturated or too close to a nearby saturated star, or epochs affected by cirrus clouds. As with our spectroscopy, epochs were identified as outliers in the light curves that deviated from the median flux by $>$ 5 times the NMAD of the light curve within each individual observing season (i.e., the NMAD was calculated using only data taken within a specific observing season, and outliers excluded from that season based on that NMAD alone rather than the entire four-year light curve). We visually inspected all of the resulting light curves to confirm that this procedure was effective.

\subsection{Light Curve Inter-Calibration and Uncertainty Corrections} 
\label{sec:intercalibration} 
To improve the precision of our continuum light curves, we placed all of the light curves from different instruments, telescopes, fields, and in different bands onto the same flux scale --- we hereafter refer to this as light-curve ``inter-calibration". This approach accounts for differences in detector properties, telescope throughputs, and properties specific to the individual telescopes. We combine both $g$- and $i$- band light curves together to increase the number of data points, assuming that the time lag between these two bands is negligible (inter-band continuum lags have been measured for some of the SDSS-RM sample by \citealt{Homayouni19}, but the measured lags are generally on the order of a week or less, which is smaller than the uncertainties for our lag measurements). 

\begin{figure}
\begin{center} 
\includegraphics[scale = 0.335, trim = 0 0 0 0, clip]{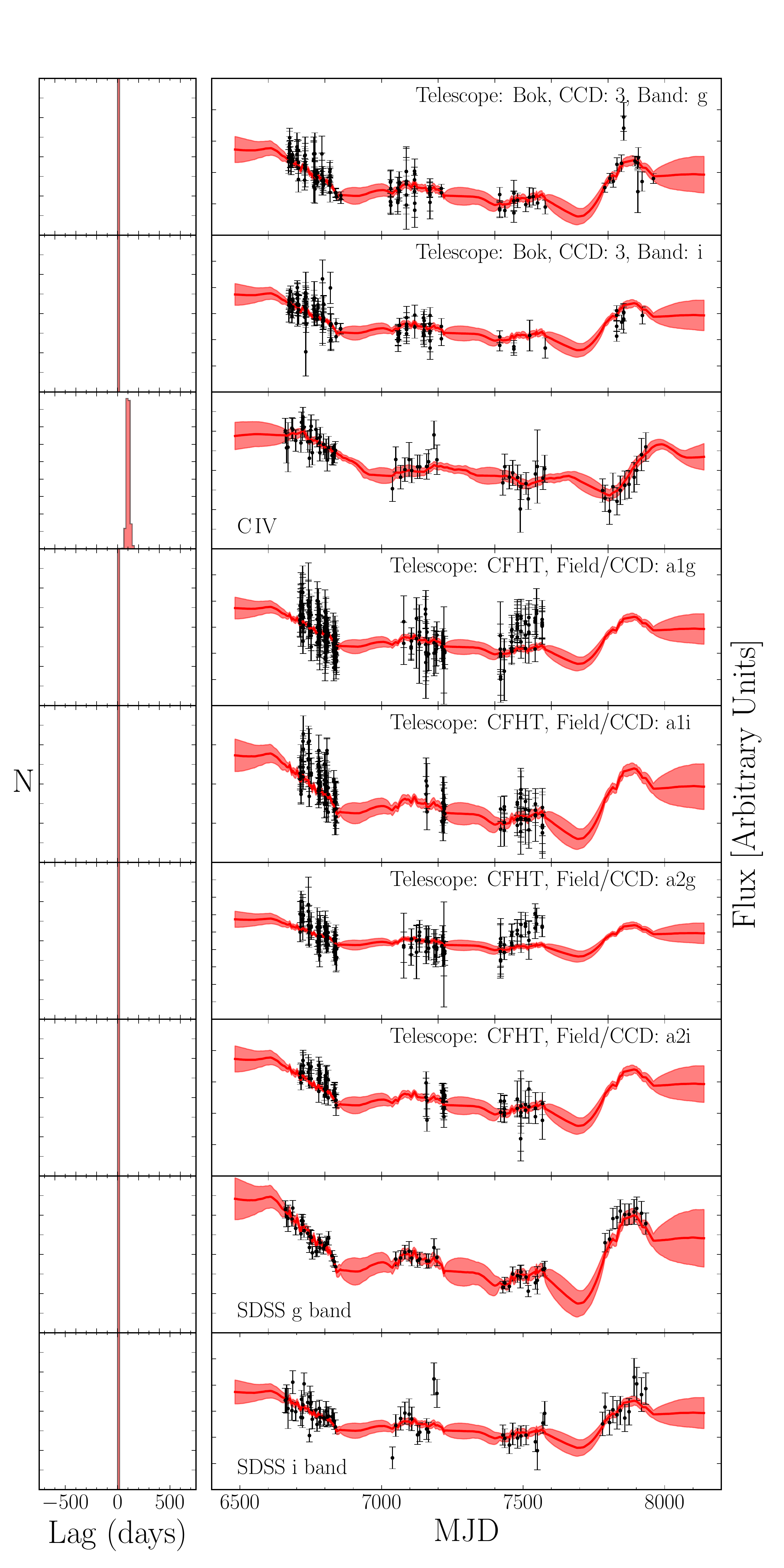}
\caption{A demonstration of the {\tt CREAM} modeling technique, using SDSS\,J141250.39+531719.6 (RM\,052) as an example. The left panels present the {\tt CREAM} posterior distributions of observed-frame time lags; the right panels show the original light curves (black filled points) with the {\tt CREAM} model fits and their uncertainty envelopes (red).} 
\label{fig:cream_example} 
\end{center} 
\end{figure} 

To combine our light curves, we use the Continuum REprocessing AGN MCMC ({\tt CREAM}) software recently developed by \cite{Starkey16}. A brief overview of this technique is provided here; see \cite{Starkey16} for details. {\tt CREAM} models the light curves using Markov Chain Monte Carlo (MCMC). The model assumes that the observed continuum emission is first emitted from a central ``lamp post" and later reprocessed by more distant gas. Each telescope/field/CCD light curve is fit to a model that includes an additive offset, scaling parameter, and transfer function (for inter-calibration purposes, we set the parameters within {\tt CREAM} such that it has a delta function response at zero lag).  After optimization via the MCMC fitting process, the rescaled $g$ and $i$ light curves are placed on the same scale as the reference light curve, and the resulting light curves are treated as a single light curve for all further analysis purposes. Figure~\ref{fig:cream_example} provides a demonstration of this procedure. 

%



The final step in our light-curve preparation considers the uncertainties in our data. The ISIS image subtraction software reports only local Poisson error contributions and neglects additional systematic uncertainties; our photometric/continuum light curve uncertainties are thus generally underestimated by a factor of a few. Similarly, PrepSpec includes only spectral uncertainties in its emission-line flux calculations. To address this, we use an additional feature of the {\tt CREAM} software that allows it to adjust the nominal error bars of the light curves. We used {\tt CREAM} to search for extra variance within the light curves and apply a multiplicative correction to the uncertainties when they are underestimated. For our quasar sample, {\tt CREAM} applied a median scale factor of 3.5 to correct the uncertainties in the continuum light curves and 2.6 for the emission-line light curves. We adopt the {\tt CREAM}-scaled light curves and their adjusted uncertainties for all further analysis. The final, inter-calibrated light curves for each source with adjusted uncertainties are provided in Table~\ref{Table:lcs}. 

 \subsection{Emission-Line Variability Contamination} 
Because we are using photometric light curves (including synthetic photometry produced from spectra) to represent the continuum light curves, we also investigate the emission lines that fall within the wavelength range covered by the $g$ and $i$-band filters. The broad emission lines are expected to be variable, and strongly variable emission lines falling within the wavelength range of the filters could have a significant impact on the photometric/continuum light curve. Significant variability contamination from the BLR would result in underestimated lag measurements, and would in effect make it more difficult to detect a lag. 

Because the lag measurements depend on the observed variability, we need to know how much of that observed variability is due to the broad emission lines instead of the continuum. 
To estimate this, we use the PrepSpec measurements of intrinsic RMS variability for the broad emission lines and continuum within the wavelength range covered by the $g$ and $i$ filters. The ``variability contamination fraction" (hereafter $f_{\rm var,BLR}$) is the sum of the variability contributions from each emission line within the FWHM of the filter: $f_{\rm var, BLR}$~=~$\sum \left( \frac{\rm RMS_{line}}{\rm RMS_{cont}} \right) \left( \frac{\rm EW_{line}}{\rm FWHM} \right)$. Here $\rm RMS_{line}$ and $\rm RMS_{cont}$ are the PrepSpec-measured fractional RMS variability of each broad emission-line and the continuum nearest the filter effective wavelength, and $\rm EW_{line}$ is the observed-frame equivalent width of the emission line measured by \citet{Shen18}. In our sources, this quantity is generally small, matching the expectation that the continuum is more variable than the emission lines (e.g. \citealp{Sun15}). We find a median variability contamination fraction of 9.1\% in the $g$ band and 1.4\% in the $i$ band in our quasar sample. In other words, the BLR contamination is negligible for most of our sources, and will be generally smaller than the measured lag uncertainties.

\section{TIME-SERIES ANALYSIS} 
\label{sec:timeseries} 
\subsection{Lag Measurements} 
We follow \cite{Grier17b}, hereafter G17, and employ three lag detection methods to analyze our sample: The {\tt JAVELIN} software (\citealt{Zu11}), traditional cross-correlation functions (CCF; e.g., \citealt{Peterson04}), and the {\tt CREAM} software (\citealt{Starkey16}). Details of each of these methods are provided in each of these works as listed; we below provide only a brief synopsis of each method. 

Our primary method for time-lag detection is the \javelin code \citep{Zu11, Zu13}, which models the light curves as an autoregressive process using a damped random walk (DRW) model, which has been demonstrated to be a good description of quasar behavior on the relevant timescales to our study (e.g., \citealt{Kelly09, Kozlowski10, Kozlowski16, Macleod10, Macleod12}). \javelin accounts for all of the likely behavior of the light curves during gaps in the light curve and applies uncertainties to the model accordingly. \javelin builds a model of both the continuum and emission-line light curves while simultaneously fitting a transfer function using Markov Chain Monte Carlo techniques. We assume that the emission-line light curves are smoothed, lagged versions of the continuum light curve, and adopt a top-hat transfer function that is parameterized by a scaling factor, width, and time delay. We allow \javelin to explore a range of observed lags from $-$750 to 750 days, which is about 60\% of the total length of our campaign. We then determine $\tau_{\rm JAV}$, the best-fit time delay, from the posterior distribution of lags that is produced by the MCMC chain, after some modifications that are described below (Section~\ref{sec:aliases}). 

Accurately modeling the light curves requires a well-constrained damping timescale ($\tau_{\rm DRW}$), and for our data, this quantity is not well fit by {\tt JAVELIN}. Prior studies using \javelin have fixed the value to be longer than the length of the observing campaign  (e.g., \citealt{Fausnaugh16a}; \citealt{Grier17b}), which effectively negates the impact of this on the time-lag measurements. Since the time baseline of the data in this work is longer than the expected damping timescales, however, we here allow this parameter to vary in {\tt JAVELIN}, but place a strong constraint on the $\tau_{\rm DRW}$ parameter. For each source, we calculate the expected $\tau_{\rm DRW}$ value based on Table~1 and Equation~7 of \cite{Macleod10}, which relates the damping timescale to the luminosity of the quasar; this expected value (typically on the order of $\sim$400--600 days for our sample) is fed into \javelin as a starting point, with small allowable uncertainties, for the MCMC step. This prevents the software from fitting un-physically small damping timescales to the data. However, the lag measurements obtained are quite insensitive to the $\tau_{\rm DRW}$ value fit by {\tt JAVELIN}; lag measurements with and without setting this constraint are almost always consistent with one another. In addition, we also fixed the width of the top-hat transfer function to 20 observed-frame days; this helps keep \javelin from fitting un-physical values when the top-hat width cannot be constrained by our data. We tested several different top-hat widths (ranging from 10 to 40 days) and the lag results came out consistent with one another regardless of the width chosen: Fixing the top-hat width produces more clean posterior lag distributions than when it is allowed to vary, but the exact value of the chosen width has negligible effect on our results. 

Historically, CCF methods have been used most frequently to measure RM lags, so we include these measurements for completeness and ease of comparison with prior results. However, we note that these methods have been reported to perform less well on datasets of similar quality to ours (e.g., G17; \citealt{Li19}); these data have more sparse time sampling and noisy light curves compared to much of the RM data for local AGN. This class of methods includes the interpolated cross correlation function (ICCF; e.g., \citealt{Peterson98}), the discrete correlation function (DCF; \citealt{Edelson88}) and $z$-transformed DCF (zDCF; \citealt{Alexander97}). We adopted the ICCF method,  as it has been used most often in previous studies, and has also been shown to perform better than the DCF in cases of low sampling (\citealt{White94}). The ICCF linearly interpolates between data points on a user-specified grid and the CCF is constructed by calculating the Pearson coefficient $r$ between the two light curves at each possible lag. The centroid of the CCF ($\tau_{\rm cent}$) is measured using points surrounding the maximum correlation coefficient $r_{\rm max}$ of the CCF. We used the {\tt PyCCF} code\footnote{The {\tt PyCCF} code is available for download at https://bitbucket.org/cgrier/python\_ccf\_code.} (\citealt{Peterson98}; \citealt{Sun18}) to perform our ICCF calculations with an interpolation grid spacing of 2 days, and again restricted our lag search to lags between $-$750 and 750 days. We calculate the best lag measurement and its uncertainties via the flux randomization/random subset sampling method, using Monte Carlo simulations, as discussed by \cite{Peterson04}. We perform 5000 realizations to obtain the cross correlation centroid distribution (CCCD) and adopt the median of the distribution; the uncertainties in either direction are set to the 68th percentile of the distribution. 

As an additional check, we report the lags measured by {\tt CREAM}, which also measures time delays while performing the inter-calibration of the light curves discussed above. \cream is similar to \javelin in many ways but it assumes a random walk model (where the Fourier transform of the time series is proportional to the square of the frequency) instead of a DRW model to interpolate the light curves (\citealt{Starkey16}). During the inter-calibration process, \cream fits a top-hat transfer function to the emission lines and reports the posterior probability distribution of lag values, from which we measure the best-fit lag ($\tau_{\rm CREAM}$). 

\subsection{Alias Identification and Removal} 
\label{sec:aliases} 
One of the hazards of obtaining RM data with regular seasonal gaps is the potential for lag-detection algorithms to prefer lags that result in the light curves being shifted into the seasonal gaps in the data; i.e., since RM lag detection algorithms interpolate or model within these gaps, they often end up associating features in the real continuum light curves with ``fake" (model, or interpolated) data in the shifted emission-line light curves. Inopportune features in the light curves can cause various lag-detection methods to latch on to incorrect lags (e.g., \citealt{Grier08}). In addition, these data (and single-season data) often possess multiple significant peaks in their lag posterior distributions that can easily be identified as aliases of a primary lag solution;  including the entire posterior distribution in the lag calculation in these cases often results in a skewed lag measurement and/or uncertainties that are unreasonably large. 

To remedy these issues, we require additional procedures beyond simply measuring the lags from the entire posterior distributions for each method. We adopt a similar procedure as G17 (see their Section~3.2), but modify the procedure to take into account the effects of seasonal gaps on the data. 
We apply a weight on the distribution of $\tau$ measurements in the posterior probability distributions --- this weight is used to search for the primary peak of the distribution and  establish a range of lags within the posterior distribution that are included in the final lag and uncertainty calculations. Our weighting procedure has two components:  
\begin{enumerate}
\item The first component takes into account the number of overlapping spectral epochs at each time delay. Applying a time lag $\tau$ to the emission-line light curve will shift the data such that fewer ``real" points will overlap. If the time lag is such that the shift results in little or no overlap between the two datasets (for example, a $\tau$ of 180 days in datasets with regular 6-month seasonal gaps), detecting that lag will be very difficult. Any potential detection of such a lag in our data has a relatively high probability of being spurious; therefore we down-weight such lags in the posterior distribution. We calculate the function $P(\tau) = [N(\tau)/N(0)]^2$, where $N(\tau)$ is the number of real emission-line data points that overlap in date ranges with the continuum data and $N(0)$ is the number of overlapping points at $\tau = 0$. Thus the weight on a lag measurement is 1 at $\tau$ = 0 and decreases each time a data point moves outside the data overlap regions. Because our data have regular 6-month annual gaps, $P(\tau)$ rises and falls as each segment of the light curve is shifted into and out of the overlapping ranges of each year of data. 
\item The second component accounts for the effect our seasonal gaps will have on our ability to detect certain lags. To characterize this phenomenon, we compute the autocorrelation function (ACF) of the continuum variations. If the ACF declines rapidly, the annual gaps will have a significant effect on our sensitivity because we are less likely to account correctly for the light-curve behavior during the gaps. In cases where the ACF declines slowly away from zero lag, it is straightforward to interpolate across the seasonal gaps, and the gaps are thus less likely to have an effect on our lag measurements.  
\end{enumerate} 

\begin{figure}
\begin{center} 
\includegraphics[scale = 0.35, trim = 0 0 0 0, clip]{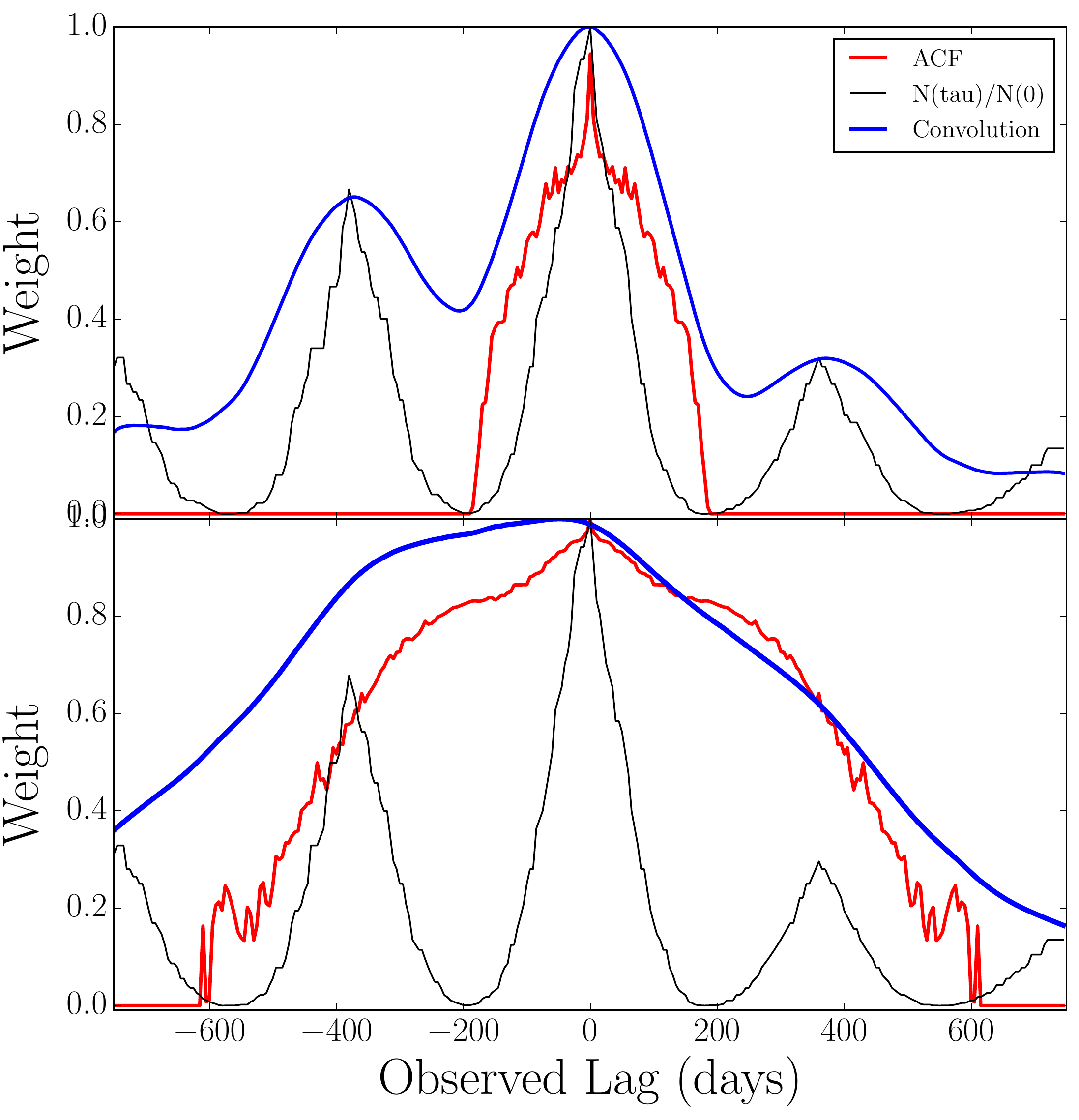}
\caption{A demonstration of the adopted weighting scheme used in our alias removal procedure. The black line indicates $P(\tau)$, the red line shows the continuum ACF (set to zero wherever it is originally less than zero), and the thick blue line is the convolution of the two, which is our final adopted weight. The top panel shows an example where the continuum ACF declines rapidly (thus making it more unlikely that we detect spurious lags within the gaps in overlapping points); the bottom panel demonstrates a case where the continuum ACF declines slowly. } 
\label{fig:ptau} 
\end{center} 
\end{figure} 

The final weight that we apply to the posterior distributions is thus a convolution of the continuum ACF and the $P(\tau)$ function, with one small adjustment: If the ACF drops below zero within our lag range, we set its value at that lag to zero before the convolution. Figure~\ref{fig:ptau} shows two examples of these functions (one with a rapidly declining ACF and one with a slowly declining ACF). We smooth the weighted posterior lag distributions (for \javelin and {\tt CREAM}, this is the posterior lag distribution, and in the case of the cross-correlation function, this is the CCCD) by a Gaussian kernel with a width of 15 days and identify the tallest peak within this smoothed distribution as the ``primary" peak. We identify local minima in the distribution to either side of the peak and adopt these minima as the minimum and maximum lags to be included in our final lag calculation. We then return to the {\it unweighted} posteriors, reject all lag samples that lie outside of the determined range, and use the remaining samples to calculate the final lag and its uncertainties. 

\begin{figure}
\begin{center} 
\includegraphics[scale = 0.185, trim = 20 55 40 0, clip]{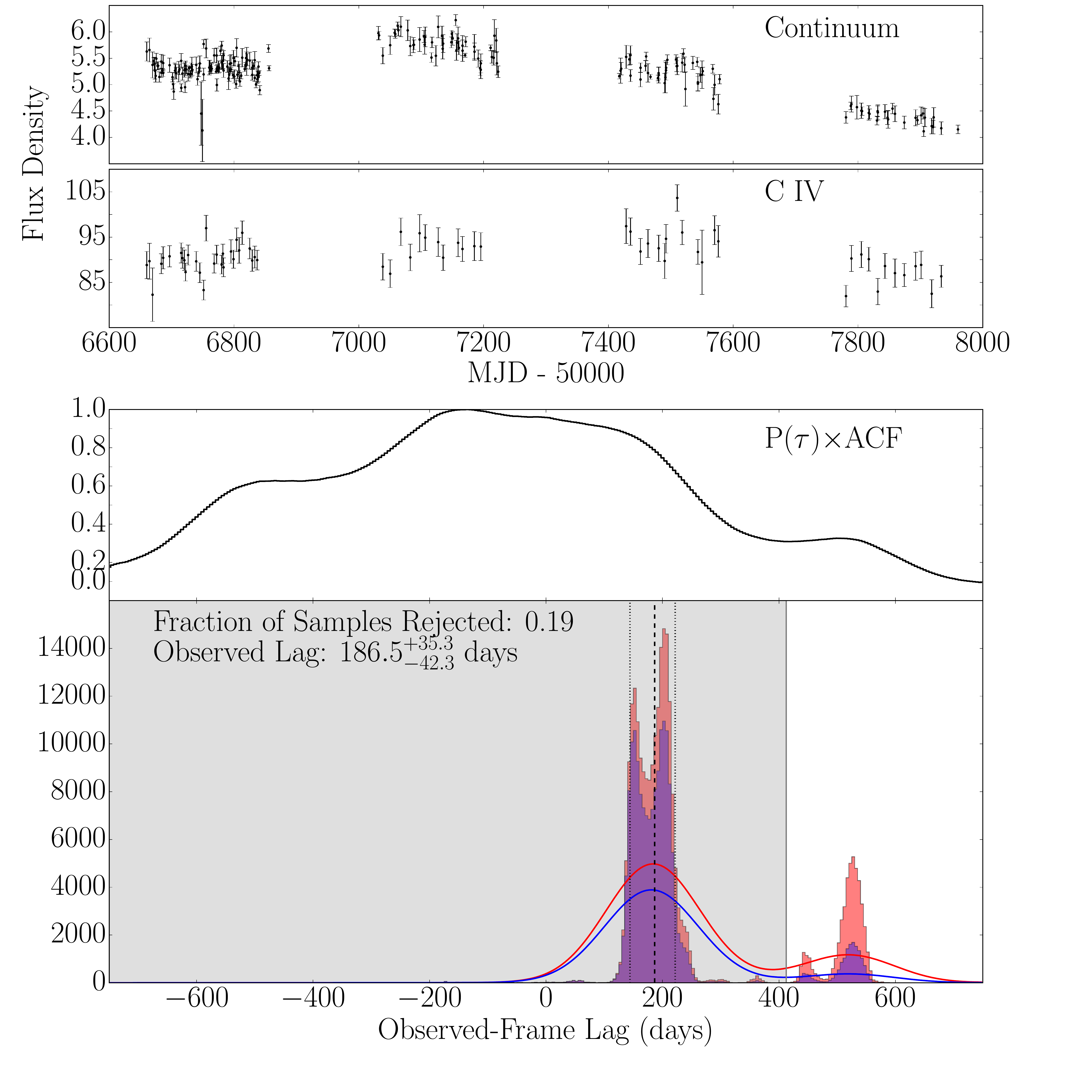}
\caption{A demonstration of our alias removal procedure. The top two panels are the light curves for RM\,119 (SDSS\,J141135.55+524814.4), with continuum flux density in units of in units of 10$^{-17}$ erg s$^{-1}$ cm$^{-2}$ \AA$^{-1}$ and integrated emission-line fluxes in units of 10$^{-17}$ erg s$^{-1}$ cm$^{-2}$. The third panel shows the adopted weighting scheme. The bottom panel shows the original \javelin posterior distribution for this object (pink histogram) and the weighted posterior distribution after applying the calculated weight (blue histogram). The solid red and blue lines indicate the smoothed posterior distribution of the original and weighted posteriors, respectively. The shaded gray region highlights the range of lags included in the final lag calculation. The dashed black vertical line indicates the measured lag and the black dotted lines show the measured uncertainties. }
\label{fig:alias_removal} 
\end{center} 
\end{figure}

The best lag is taken to be the median of the distribution, with the uncertainty in either direction calculated using samples within the 68th percentile of the distribution. Figure~\ref{fig:alias_removal} provides a demonstration of this procedure for one of the quasars in our sample. We tested this alias removal approach with mock light curves (with known lags) that mimic the SDSS-RM data, and found that this approach is very efficient in removing alias lags (\citealt{Li19}). 

\subsection{Lag-Significance Criteria} 
\label{sec:significance} 
While our alias-removal procedure above mitigates the problem of lag aliases and seasonal gaps, these methods are not foolproof. The fact remains that in some cases, the lags are just not well measured, despite the models reporting their best solutions. Following G17, we thus impose a number of additional criteria on our measurements for a lag to be considered a significant detection: 

\begin{enumerate} 
\item The lag can be positive or negative, but must be inconsistent with zero at 1$\sigma$ significance.
\item Less than half of the posterior lag samples can be removed by our alias-removal procedure described in Section~\ref{sec:aliases}. If a larger fraction of samples are eliminated by this procedure, this indicates that most of the samples lay outside of the primary peak that we identified, suggesting that we lack a solid measurement of $\tau$. 
\item The behavior of the light curves must be well correlated at or near the measured lag, as characterized by the Pearson correlation coefficient $r$ measured by the ICCF. We include only measurements of quasars for which $r$ reaches a value greater than 0.5 within $\pm$1$\sigma$ of the reported lag (see below for a discussion of how this threshold was chosen). 
\item When selecting our quasar sample, we required that the emission-line light curves showed some variability (see Section~\ref{sec:sample}). However, after merging the light curves and adjusting the uncertainties of the light curves, some sources are no longer significantly variable. We thus require that both the continuum and emission lines are still considered significantly variable after the inter-calibration process. To quantify this variability, we follow G17 and measure the RMS variability signal-to-noise ratios (SNR) in the merged/adjusted light curves. We require that the continuum and emission-line RMS variability SNR (SNR$_{\rm con}$ and SNR$_{\rm line}$) are greater than 6.5 and 2.0, respectively. This criterion effectively eliminates cases where the light curves are consistent with little-to-no real variability, which can result in the lag detection methods latching on to monotonic trends or spurious correlations between noisy light curves. Roughly 20\% of the 349 quasars do not meet this criterion for SNR$_{\rm line}$ --- however, all but two of these sources also fail additional criteria and would thus not have been selected as significant lags regardless. 
\end{enumerate} 

\begin{figure}
\begin{center} 
\includegraphics[scale = 0.45, trim = 0 0 0 0, clip]{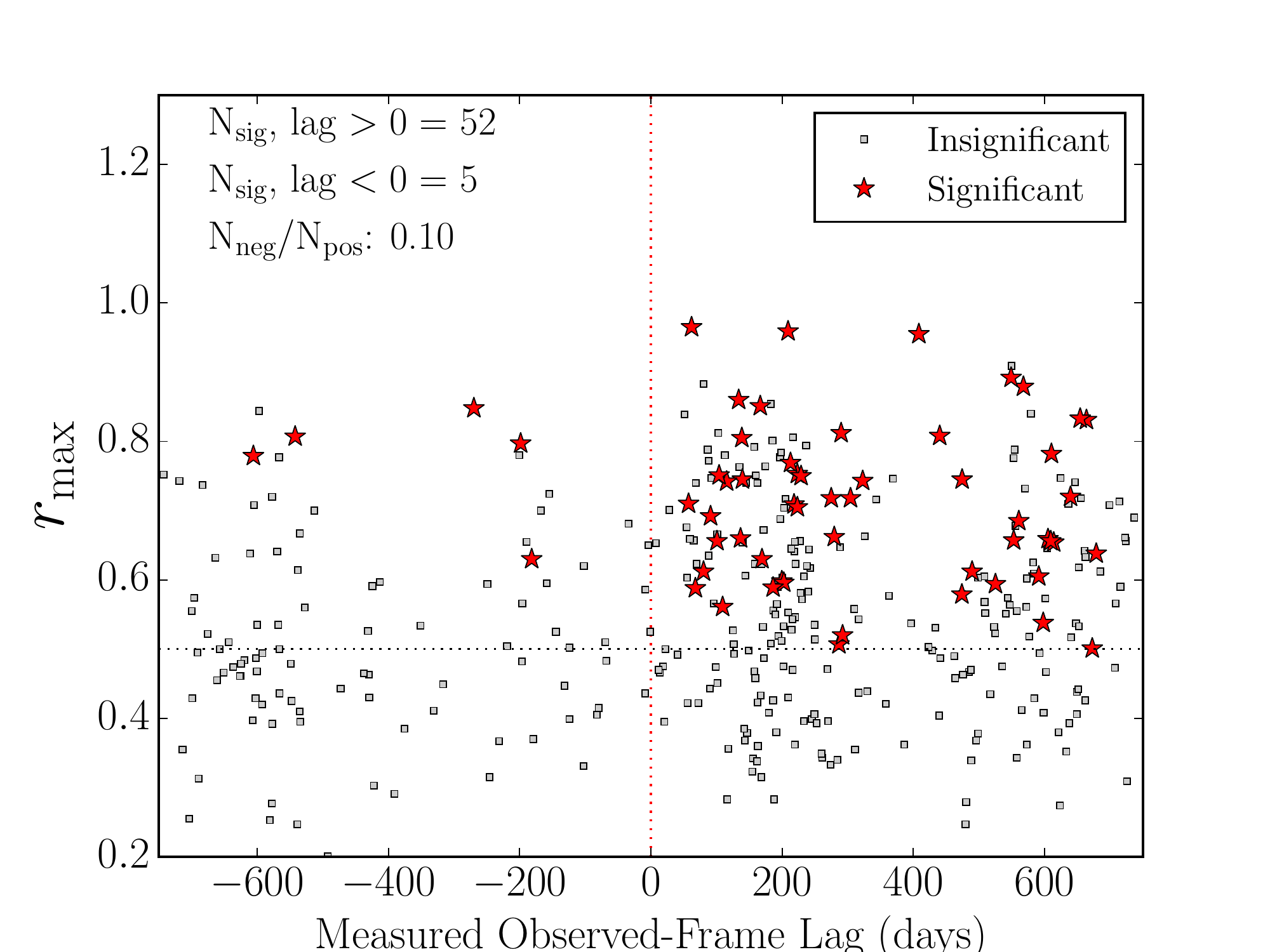}
\caption{The measured time lag vs. $r_{\rm max}$ for all quasars in our sample. Those measurements that do not meet the criteria for significant lags are shown as gray points; those that meet all of the significance criteria are represented by red stars. The vertical dotted red line indicates a lag of zero to guide the eye, and the horizontal dotted black line indicates the threshold of $r_{\rm max}$ = 0.5 used to select our significant lag sample. }
\label{fig:nielplot} 
\end{center} 
\end{figure}

To determine the thresholds for $r_{\rm max}$, SNR$_{\rm con}$, and SNR$_{\rm line}$, we utilize a positive/negative false-positive test as implemented by \cite{Shen16}, G17, and \cite{Li19}. We assume that there is no physical reason to measure a negative lag; if all lag measurements were the result of spurious correlations rather than physical processes, we would expect to measure equal numbers of positive and negative lags in our sample\footnote{We ran simulations to confirm that asymmetric sampling in our data does not bias our results toward either positive or negative lags.}. We can thus use the number of negative lag measurements to estimate the rate of false-positive detections at positive lags in our sample. We define the ``false-positive rate" as the ratio of negative lags to positive lags. Even including all of our lowest-quality measurements, we see a strong preference for positive lags: Without imposing any selection criteria at all, we have 254 positive measurements and 95 negative measurements (see Figure~\ref{fig:nielplot}), which indicates a false-positive rate of 37\%. We provide all 349 measurements as well as the quantities by which we measure their significance in the Appendix in Table~\ref{Table:appendixtbl}. 

We choose the thresholds for our selection criteria described above to lower our false-positive rate to an acceptable level while maximizing the number of positive lag detections. We choose a maximum acceptable false-positive rate of 10\%. Figure~\ref{fig:nielplot} shows the resulting distribution of lags for both those deemed ``insignificant" and those passing our selection criteria. By down-selecting the sample to a false-positive rate of 10\%, we exclude many true lags --- based on the false-positive rate without imposing our additional constraints, we expect that there is on the order of $\sim$100 additional measurable lags in our sample. Such lags may be recoverable with additional years of data. 


We adopt \javelin as our primary lag-detection method and therefore require that all of our significance criteria are satisfied specifically for the \javelin measurements. This results in \numlags positive lag detections and \numneg negative measurements in our full ``primary" sample of lag detections. 

\begin{figure}
\begin{center} 
\includegraphics[scale = 0.3, trim = 0 0 0 0, clip]{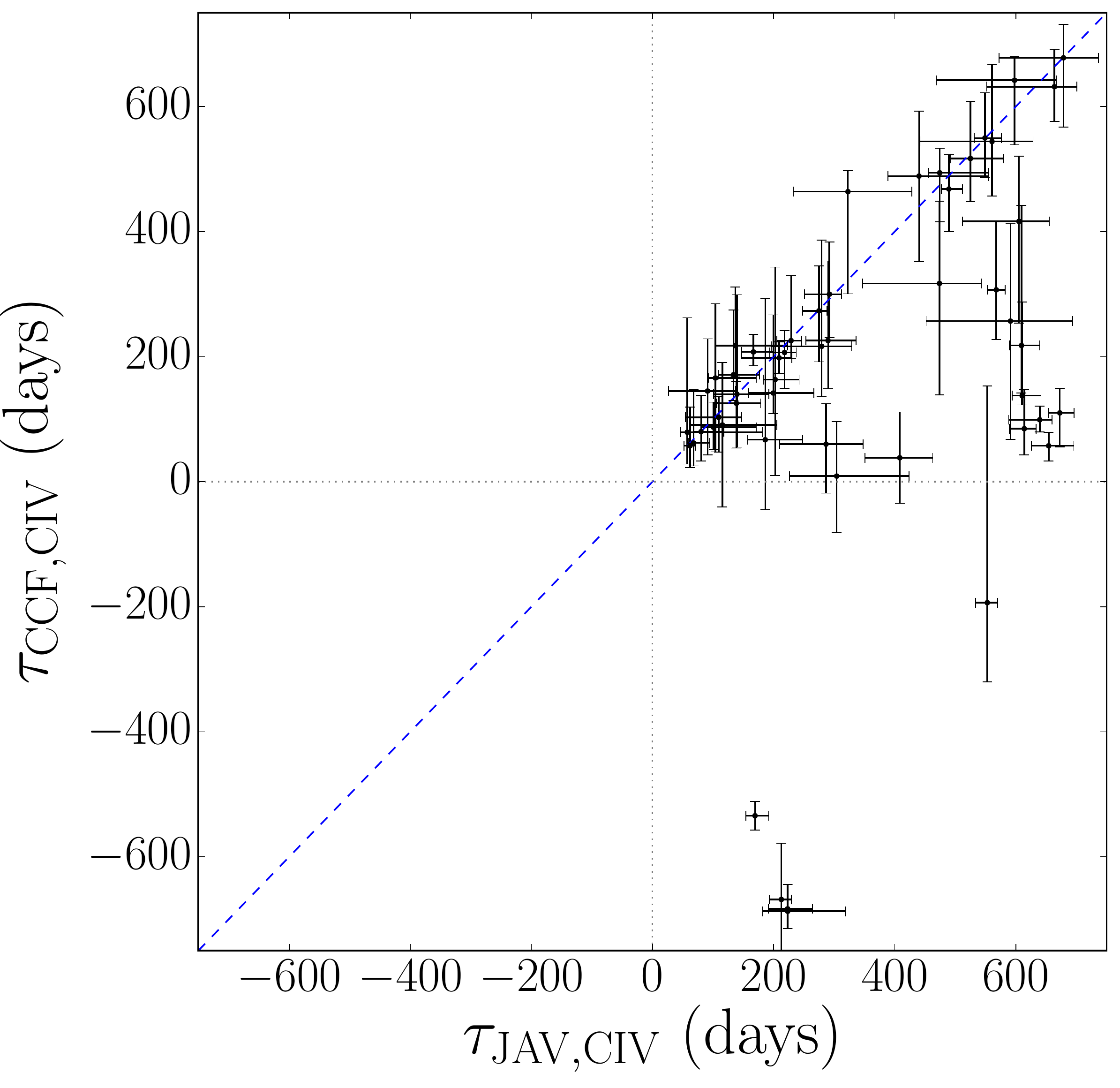}
\includegraphics[scale = 0.3, trim = 0 0 0 0, clip]{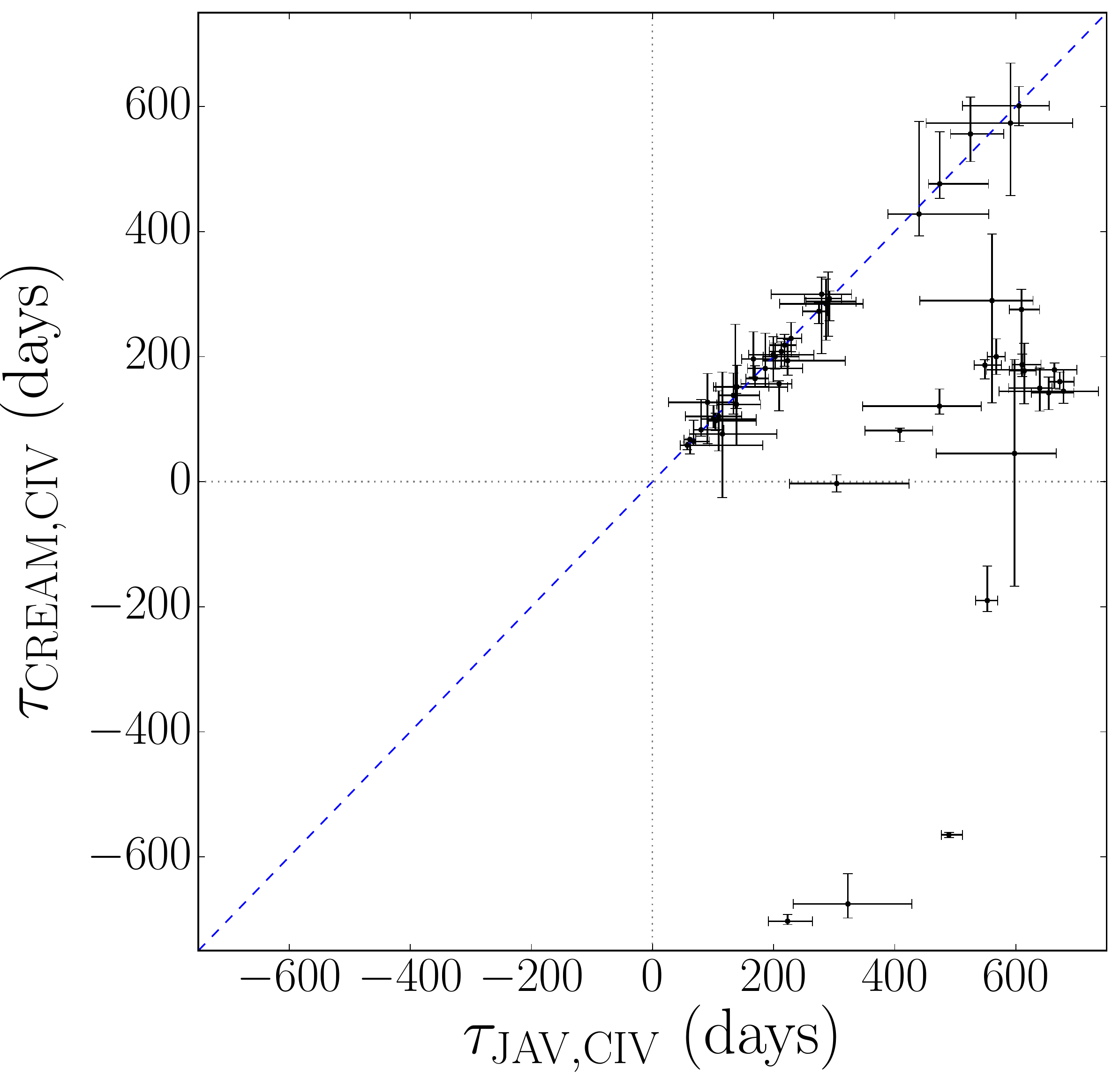}
\caption{A comparison of the observed-frame lag measurements made using the different detection methods for our \numlags positive lag detections. The top panel shows the lags measured by the ICCF vs. the \javelin measurements, and the bottom panel presents lags measured by \cream vs. the \javelin measurements. }
\label{fig:methodcomp} 
\end{center} 
\end{figure}

For comparison purposes, we apply these selection criteria separately to the lags measured with all three methods. In about 2/3 of our lag measurements, the resulting lags from all three methods are consistent with one another (see Figure~\ref{fig:methodcomp}). As reported by G17 and others (e.g., \citealt{Li19}), the ICCF generally produces larger uncertainties than \javelin and {\tt CREAM}, and the ICCF is less sensitive than JAVELIN to lag detection with light curve qualities similar to SDSS-RM (\citealt{Li19}). There has been some discussion in the literature (e.g., \citealt{Edelson19}) regarding the uncertainties reported by {\tt JAVELIN}; i.e., it has been suggested that \javelin uncertainties are underestimated. Simulations are under way to resolve this open issue, but in the meantime, we note that 41 out of \numlags of our significant lags were also formally detected by the ICCF method, which is widely suspected to {\it overestimate} the lag uncertainties, and while we chose 1$\sigma$ as our detection threshold, all but four of them are $>2\sigma$ detections. Our detections are thus robust against the possibility that the uncertainties reported by \javelin are underestimated to within a reasonable extent. 

For about a third of our measurements, the ICCF or \cream software reported different alias lags than {\tt JAVELIN}; in these cases, a different primary peak was identified, resulting in lag disagreements. In all of these cases, we see the same aliases present for all three methods, but their strengths vary, causing different lags to be preferred by different methods. 
We have visually inspected all of the cases where the three measurement methods disagree and confirm that the peaks identified by \javelin are reliable in most cases; cases where the \javelin lags appear to be incorrect are taken into account with our lag measurement quality ratings (discussed below in Section~\ref{sec:goldsample}). 

\subsection{Lag Measurement Quality and the ``Gold" Sample} 
\label{sec:goldsample} 
\subsubsection{Quality Ratings} 
Though our false-positive test (Figure~\ref{fig:nielplot}) indicates that the majority of our lag measurements are robust,  because our lag-selection procedure uses statistical arguments and we apply our criteria to achieve a false-positive rate of 10\%, it is statistically likely that the lag sample presented here contains false detections. There is a subset of our lag detections that have characteristics indicating that they are more likely to be real than others; we thus follow G17 and assign quality ratings to each of our measurements in order to help readers assess the results. We use a scale of 1 to 5, with 1 representing the poorest-quality measurements and 5 representing the highest-quality measurements. We took into account a variety of criteria when assigning these quality ratings: 

\begin{enumerate} 
\item{There are visible variability features in the continuum light curve that appear also in the emission-line light curve; i.e., it is possible to pick out a ``lag" between the two light curves by eye.} 
\item{There is clearly defined structure corresponding to the \civ\ emission line in the RMS line profile (see Figure~\ref{fig:meanrms} in the Appendix).} 
\item{The model fits from \javelin and \cream match the light-curve data well, and there is general agreement in the models between the two methods.} 
\item{The ICCF has a clear, well defined peak on or around the measured lag.} 
\item{There is general agreement between the three different methods used.} 
\item{Unimodality of the posterior lag distribution: If there are several other peaks with comparable strengths to the peak that was determined to be the primary peak, this reduces our confidence in a lag measurement.} 

\end{enumerate} 

We include these quality ratings, assigned by the first author of this work, in Table~\ref{Table:lags}. In addition, we place all of the measurements with quality ratings of 4 and 5 into a ``gold sample" of lag measurements that represent our highest-confidence individual measurements. Our gold sample includes \numgold sources. We note that the criteria used to rate the lag measurements are subjective and based primarily on our prior experience with RM measurements. Thus, our gold sample is not statistically meaningful and should not be interpreted as such.

\subsubsection{Broad Absorption-Line Contamination} 
\label{sec:BALs}
 Because we are focused on the \civ\ region of the spectrum, we must also consider the possible presence of broad and narrow absorption features. PrepSpec does not currently fit absorption profiles; for narrow absorption lines (NALs), it generally has little issue interpolating across the absorption line. This will not affect our variability measurements, though the actual integrated emission-line flux values may be offset from the true values. However, broad absorption lines (BALs) are a potential issue. When there are BALs superimposed on the \civ\ emission line, PrepSpec is often unable to correctly interpolate over the feature and the result is that the BAL is fit as part of the continuum or emission line. 

BALs are known to be variable, and they may vary simultaneously with the continuum (e.g., \citealt{Barlow93}; \citealt{Lundgren07}; \citealt{Wang15}). This may cause a light curve to be biased toward zero (or at least shorter) lags. Though studies have generally avoided BALs that are superimposed onto emission lines due to difficulties in disentangling the two, detached BALs that are at lower velocities have been reported to be less likely to vary than those at higher velocities (e.g., \citealt{Capellupo11}; \citealt{Filizak13, Filizak14}).  Low-velocity troughs are also often highly saturated and thus have depths that are unaffected by quasar variability. Assuming that these trends hold true for BALs at low enough velocities to overlap with the emission lines, we can expect any effect on lag measurements to be minimal in our sample. 

There is a second potential issue, however: An improper fit to the \civ\ line profile due to the presence of a BAL will result in incorrect line-width measurements, both for the mean line profile and for the RMS line profile (see Figure~\ref{fig:meanrms} in the Appendix for examples). This will in turn affect our \mbh measurements (see Section~\ref{sec:mbh}), which rely on accurate characterization of the line widths. Thus, \mbh measurements for objects whose RMS profile is significantly impacted by the fit around the BAL are potentially suspect, though we note that the uncertainties in the \mbh measurements are large, and  the BALs may not cause deviations outside of the measurement uncertainties. 

There are ten quasars in our lag-detected sample that have significant BAL components that overlap with the \civ\ emission line that appear to have affected the PrepSpec fits (See Figure~\ref{fig:meanrms}). All but one of these sources (RM 722) show a lower RMS within the BAL trough than in the surrounding spectrum; this is indicative of the BALs being much less variable than the emission lines, which further increases our confidence that our lag measurements are not significantly affected. We thus choose to leave these quasars in our sample. However, in Tables~\ref{Table:lags} and ~\ref{Table:mbh} and all subsequent figures, we flag all of the quasars that include significant BAL contamination to indicate the higher uncertainty and potential for error in their measurements. In addition, the severity of the BAL contamination was taken into consideration when assigning the quality ratings that are reported in Table~\ref{Table:lags}. As the figures and tables demonstrate, these sources do not deviate systematically from the positions of the non-BAL quasars, which suggests that any effect of the BALs on our results is minimal. 
\vspace{5mm}

\section{RESULTS AND DISCUSSION}
\label{sec:discussion} 

\subsection{Lag Results} 
\label{sec:lags} 
\begin{figure*}
\begin{center} 
\includegraphics[scale = 0.35, trim = 0 40 0 0, clip]{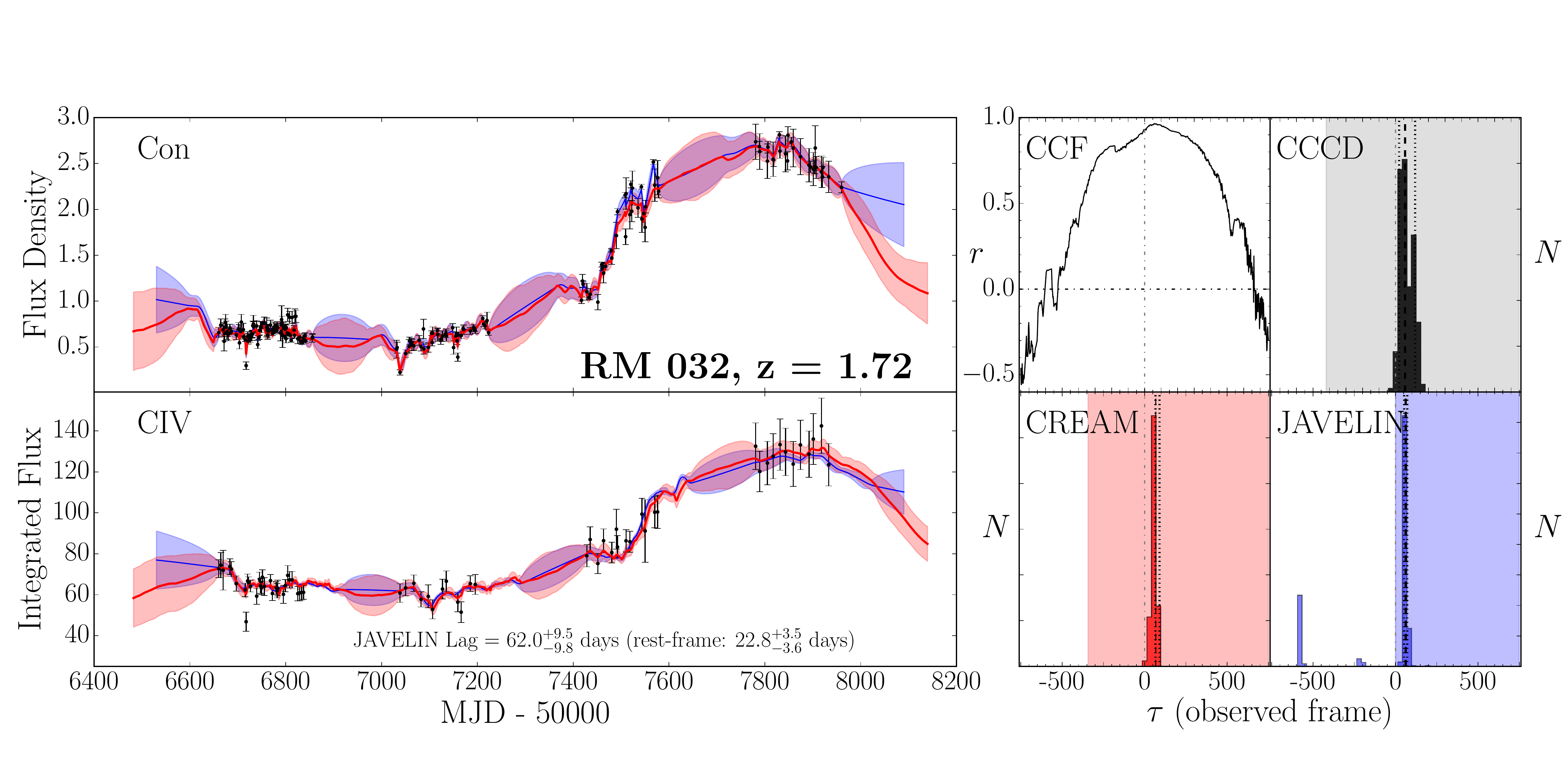}
\caption{Light curves and posterior distributions for the quasars with significant \civ\ lags in our primary lag sample. The two left panels show the continuum (top) and \civ \ (bottom) light curves: Black points are the data, blue lines show the \javelin model fit to the data (with the uncertainties shown as a blue envelope), and red lines show the \cream model fit (with uncertainties shown as a red/pink envelope). For visualization purposes, data points within a single night were combined using a weighted average. Continuum flux density is provided in units of in units of 10$^{-17}$ erg s$^{-1}$ cm$^{-2}$ \AA$^{-1}$ and integrated emission-line fluxes in units of 10$^{-17}$ erg s$^{-1}$ cm$^{-2}$. The right panels indicate the time series analysis results: The top panels show the CCF (left) and CCCD (right), and the bottom panels show the \cream and \javelin posterior distributions (left and right, respectively). The measured lag and its uncertainties are indicated as dashed and dotted lines, and the shaded regions indicate the range of lags considered in the final measurement, as per our alias rejection procedure. Figures for all of our significant lag detections are provided in the online version of the article. Sources that are affected by BALs (see Section~\ref{sec:BALs}) are flagged with red ``BAL" text in the bottom-left panel. }
\label{fig:posterior_plots} 
\end{center} 
\end{figure*}

We identify significant positive lags in \numlags quasars in our primary sample. Of these, \numgold are deemed to be high-confidence lags that constitute our ``gold sample" of lag detections. All \numlags positive lag measurements that constitute our sample are listed in Table~\ref{Table:lags}. 
Light curves, model fits, and posterior lag distributions are shown for all of our positive lag detections in Figure~\ref{fig:posterior_plots}. 

\subsection{The \civ \ Radius-Luminosity Relation}  
\label{sec:rl}
To place our measurements on the \civ\ \radlum\ relationship, we measure log$\lambda L_{\lambda1350}$, the luminosity at 1350\,\AA, from the PrepSpec model fits. In our 10 lowest-redshift sources, 1350 \AA\ was not covered by the spectrum; in these sources we measure the luminosity at 1700\,\AA\ and convert the values to  
log$\lambda L_{\lambda1350}$ by multiplying $L_{\lambda1700}$ by factor of 1.09, which was computed from the mean quasar luminosities reported in Table~3 of \cite{Richards06}. The uncertainties on the luminosity measurements provided in Table~\ref{Table:sample} include only statistical uncertainties; due to the variability of the quasars, the actual uncertainties in the average quasar luminosities are somewhat higher. To quantify this additional source of uncertainty, we calculate the standard deviation in the flux at 1350\,\AA\ for our targets and add it to the statistical uncertainties.  

\begin{figure*}
\begin{center} 
\includegraphics[scale = 0.7, trim = 0 0 0 0, clip]{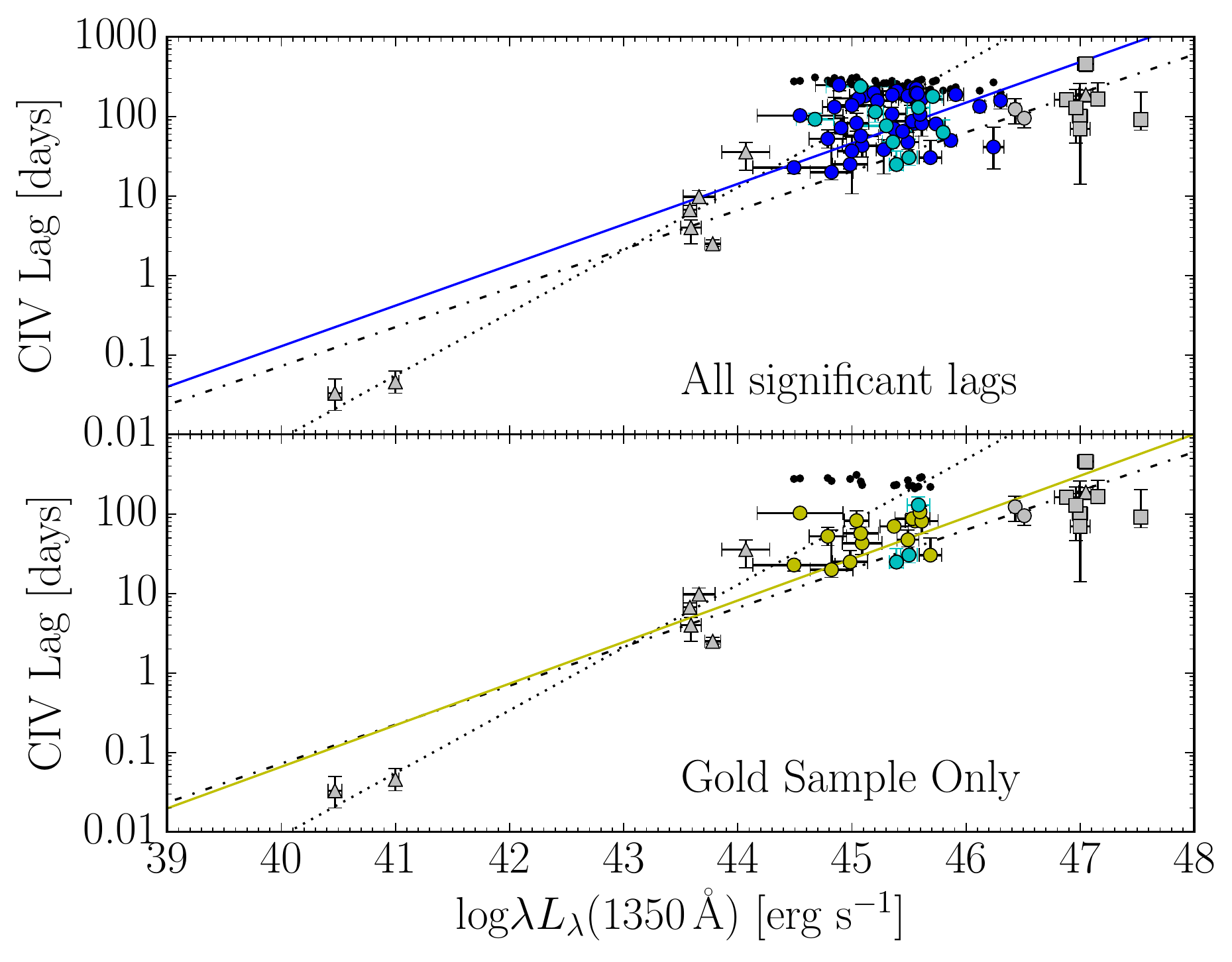}
\caption{The CIV \radlum\ relation. Gray solid triangles represent measurements from \cite{Peterson04}, who reanalyzed \civ\ data from \cite{Reichert94},  \citealt{Rodriguezpascual97}, \cite{Korista95}, \cite{OBrien98}, and \cite{Wanders97}, and additional measurements from \cite{Peterson05}, and \cite{Kaspi07}. Gray squares represent data from \cite{Lira18}, and gray circles indicate the two measurements from \cite{Hoormann19}. The dashed black lines show the best-fit line from \cite{Peterson05}, while the dashed-dotted black lines indicate the most recent best-fit line from \cite{Hoormann19}. In the top panel, the blue filled circles represent all of our significant lag measurements  and the blue solid line indicates the measured \radlum\ relation from the entire sample. In the bottom panel, the yellow filled circles represent only our measurements that we placed in the gold sample, and the yellow solid line represents the measured \radlum\ relation while including only gold-sample measurements. Cyan filled circles indicate sources that are affected by BALs (see Section~\ref{sec:BALs}). Black solid dots represent a 750-day observed-frame lag cutoff at the redshift of each of our sources; i.e., each of our measurements has a corresponding black dot that shows the longest lag we could have detected with our campaign at that quasar's redshift (see text in Section~\ref{sec:rl}).} 
\label{fig:radlum} 
\end{center} 
\end{figure*}

Figure~\ref{fig:radlum} shows the location of our sources on the \radlum\ relation. Previous recent measurements of the relation included only $\sim$15 sources (\citealt{Lira18}; \citealt{Hoormann19}); our measurements raise this number to 67. In addition, our measurements span two orders of magnitude in luminosity in a region that was previously unpopulated on the \civ\ \radlum\ relation. In general, our measurements lie fairly close to their expected locations based on previously-measured \radlum\ relations. 

We use the {\tt LINMIX} procedure described by \cite{Kelly07b} to fit a new relationship including our new measurements, which includes a measurement of the intrinsic scatter of the relation. 
We fit the relation in the form 
\begin{equation} 
{\rm log}\ R_{\rm BLR}\ ({\rm light-days}) = a + b \times {\rm log}\frac{\lambda L_{\lambda}(1350\,\AA)}{10^{44} {\rm erg s}^{-1}} + \epsilon
\end{equation} 

where $\epsilon$ is the intrinsic random scatter of the relation. The resulting line fits are shown in Figure~\ref{fig:radlum}. Including our entire sample of significant lags, we measure a slope of 0.51 $\pm$ 0.05, an intercept of 1.15 $\pm$ 0.08, and an intrinsic scatter of 0.15 $\pm$0.03. Our measured slope is consistent with the most recent measurements by \cite{Lira18} and \cite{Hoormann19}, though somewhat shallower than earlier measurements by \cite{Peterson05} and \cite{Kaspi07}. In addition, our measured intercept is larger than that measured by \cite{Hoormann19}. Previous studies used a variety of methods to measure the line fit; for comparison purposes, we also fit our relation using the Bivariate Correlated Errors and Intrinsic Scatter (BCES) method (\citealt{Akritas96}, implemented with the publicly available code of \citealt{Nemmen12}). Results from the BCES method are consistent with those using {\tt LINMIX}\footnote{Using the BCES method, we measure a slope of 0.49 $\pm$ 0.08 and an intercept of 1.15 $\pm$ 0.13.}. 

Because our full sample likely includes some false-positive measurements, we also fit the relation while including only the measurements in our gold sample (see Section~\ref{sec:goldsample}) and the previously reported measurements. We measure a slope of 0.52~$\pm$~0.04, an intercept of 0.92~$\pm$~0.08, and an intrinsic scatter of 0.11$\pm$0.04. The slope is consistent with that measured using our full sample (and with that measured by \citealt{Hoormann19} and \citealt{Lira18}). 

Our measurements occupy a previously empty region in luminosity space on the \civ\ \radlum\ relation. Additional measurements at the high-luminosity end of the relation will be possible with additional data; the SDSS-RM program will eventually include 10 years of monitoring, which will allow the detection of longer time lags in more luminous quasars and a better understanding of the intrinsic scatter at these luminosities. However, the lack of measurements at the low-luminosity end of the relation is still somewhat problematic --- the only two measurements in sources with luminosities below 10$^{43}$ erg~s$^{-1}$ lie below our measured relation. It could be that these measurements are consistent with the relation to within the expected intrinsic scatter; additionally, there may be an intrinsic difference in the accretion and/or line-emission region between low-luminosity sources and the high-luminosity quasars that populate much of the relation. Future RM experiments in the UV focused on local, low-luminosity AGN would be greatly beneficial in determining if this is the case and in more concretely constraining the slope of this relation.  

Finally, we caution that the fit of the \radlum\ relation here (and in earlier work) does not take into account selection effects in the sample. For example, our study is unable to detect lags than $\pm$750 observed-frame days, and so may be biased to short lags at the luminous end of the \radlum\ relation. Figure~\ref{fig:radlum} shows that the majority of our measurements fall well below the rest-frame equivalent of our 750-day detection threshold (for example, 750 observed-frame days translates to 250 rest-frame days for a quasar at a redshift of 2). This suggests that our lag measurements are unlikely to be biased low due to the 750-day observed-frame lag detection limit; if this was the case, we would expect many of our measurements to lie close to the upper detection limit. However, at the high-luminosity end of our sample (log$\lambda_{L\lambda}$ $>$ 45.5), the expected rest-frame time lags based on the \radlum\ relation are on par with the rest-frame time lag threshold for the range of redshifts of our sample. It is thus likely that we are missing some of the lags at the high-luminosity end due to their likely scatter above the relation (and thus above our detection threshold). 

A more detailed quantification of the selection effects on the measured \radlum\ relation is beyond the scope of this paper, and will be investigated with future SDSS-RM work that specifically focuses on the \radlum\ relation using simulations similar to those performed by \cite{Li19} and Fonseca et al. (in preparation). For this reason, the preliminary \civ\ \radlum\ relation presented here is primarily used as a sanity check on the bulk reliability of our \civ\ lags, and we do not recommend its usage for other applications (e.g., SE masses).

\subsection{Black-Hole Mass Measurements} 
\label{sec:mbh} 
For each quasar, we measure \mbh with Equation~\ref{eq:mbh} using our adopted rest-frame time lags from \javelin and line widths measured by PrepSpec during the fitting process. We adopt $\sigma_{\rm line, rms}$ as our line width measurement to compute the virial product, as past studies (e.g. \citealt{Peterson11}) have suggested that $\sigma_{\rm line, rms}$ is a less biased estimator for \mbh than the FWHM for a number of reasons. For example, the relationship between FWHM and $\sigma_{\rm line}$ is not linear, which can cause the underestimation of low masses and the overestimation of high masses when FWHM is used. In addition, FWHM measurements can often be significantly affected by narrow line components (see, e.g.,  \citealt{Wang19} for a recent discussion on this topic). However, this issue is still in contention, so we include several different characterizations of line width in Table~\ref{Table:mbh}.  We again note that some of our objects have significant BAL contamination that has affected the PrepSpec fits (see Section~\ref{sec:BALs}); we flag such cases in Table~\ref{Table:mbh} and caution that \mbh measurements for these sources may be inaccurate.

\begin{figure}
\begin{center} 
\includegraphics[scale = 0.35, trim = 0 0 0 0, clip]{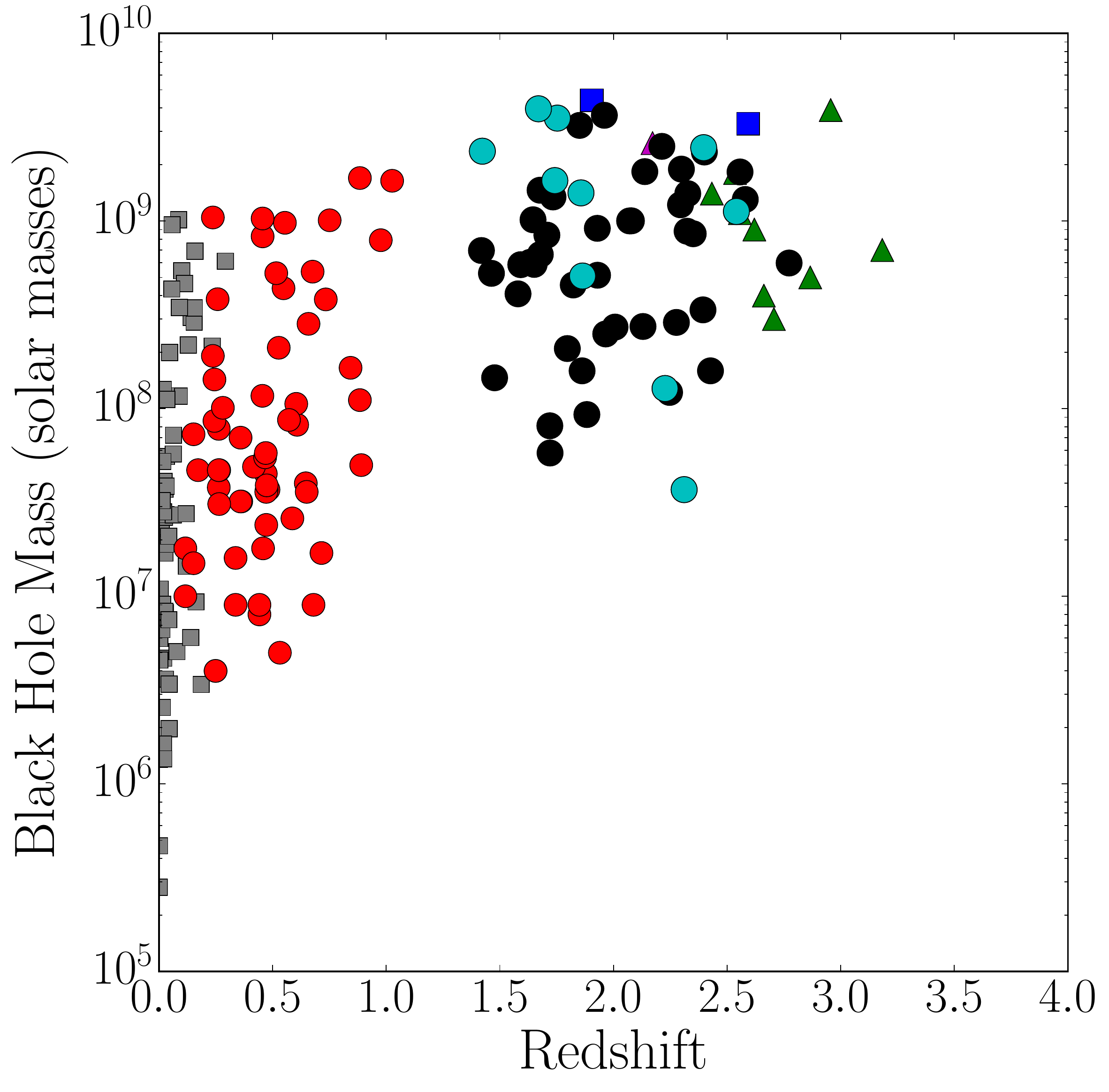}
\caption{Black hole mass vs. redshift for reverberation-mapped AGN. Gray squares represent \Hbeta\ RM measurements made prior to the SDSS-RM program (\citealt{Bentz15}, with additions from \citealt{Du16a}). Red circles indicate SDSS-RM measurements made using the \Hbeta \ emission line by G17. Blue solid squares are \civ \ measurements by \cite{Hoormann19}, solid green triangles are \civ\ measurements by \cite{Lira18}, the solid magenta triangle is from \cite{Kaspi07}, and solid black circles represent \civ\ measurements from this work. Cyan circles indicate sources from this work that are affected by BALs (see Section~\ref{sec:BALs}).}
\label{fig:mbh_vs_z} 
\end{center} 
\end{figure}

When calculating the uncertainties in the virial products, we follow G17 and add a 0.16 dex uncertainty in quadrature to the statistical uncertainties (which are calculated via standard propagation) to account for systematic uncertainties that have not been taken into account, following the 0.16-dex standard deviation among the many different mass determinations of NGC\,5548 (\citealt{Fausnaugh17}). To convert the virial products into $M_{\rm BH}$, we adopt $f = 4.47$ (\citealt{Woo15}). All virial products and \mbh measurements are provided in Table~\ref{Table:mbh}. Our \mbh \ measurements range from about 10$^8$ to 10$^{10}$ solar masses, and are among the most massive SMBHs to have RM mass measurements (see Figure~\ref{fig:mbh_vs_z}). 

\begin{figure}
\begin{center} 
\includegraphics[scale = 0.35, trim = 0 0 0 0, clip]{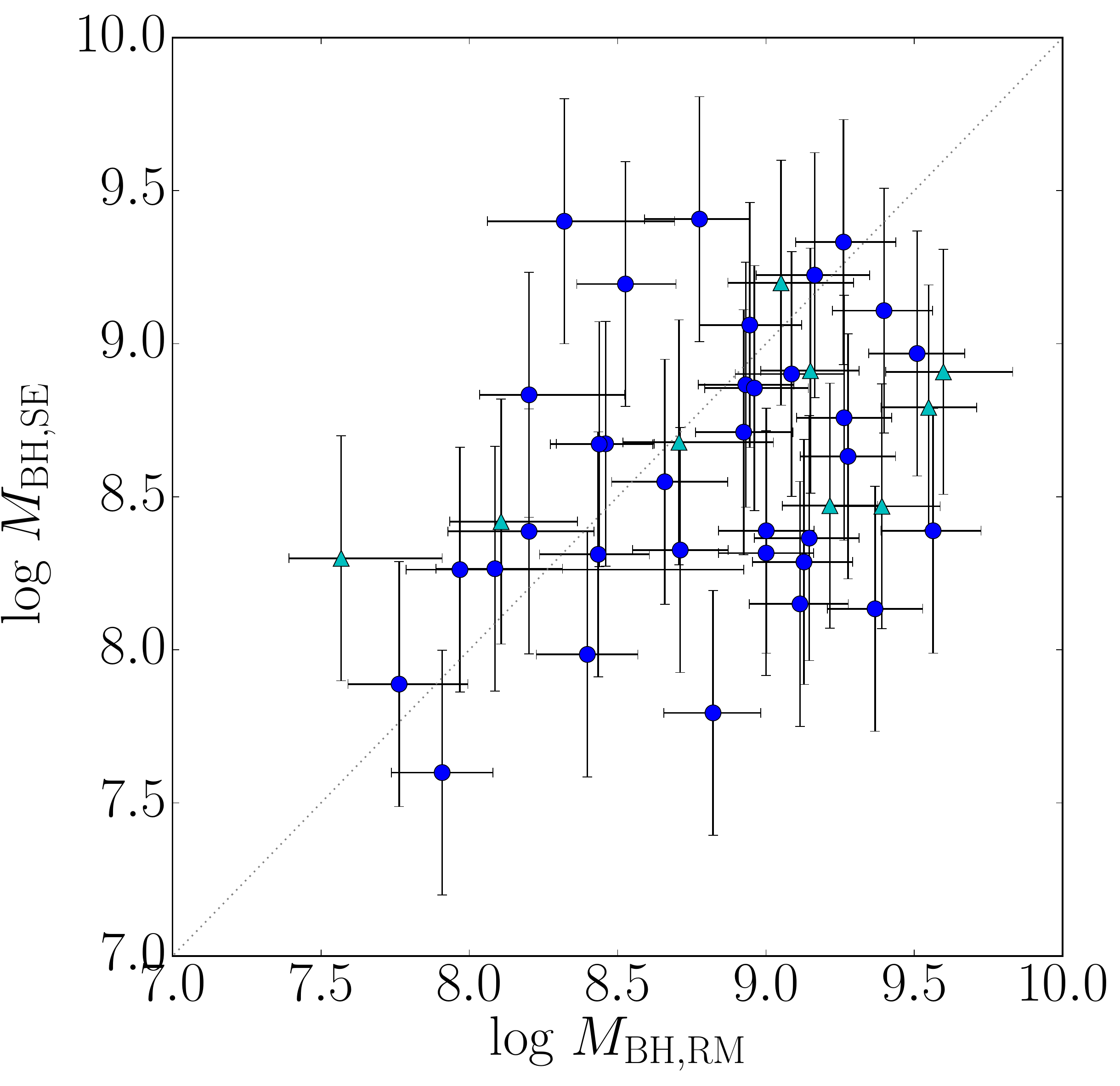}
\caption{Single-epoch \mbh estimates from \cite{Shen18} compared to our new RM measurements. Filled blue circles represent sources without BAL contamination and filled cyan triangles indicate sources with BALs (see Section~\ref{sec:BALs}). The SE values were computed using estimators from \cite{Vestergaard06}. We have increased the statistical uncertainties on the SE masses by 0.4 dex (see Section~\ref{sec:mbh}) to account for systematic uncertainties. The gray dotted line shows a 1:1 ratio.} 
\label{fig:mbhSEvsRM} 
\end{center} 
\end{figure}

Figure~\ref{fig:mbhSEvsRM} compares our RM \mbh measurements with SE \mbh estimates from \cite{Shen18}. We add systematic uncertainties of 0.4 dex to the SE measurements to the measurement uncertainties in the \cite{Shen18} values (e.g., \citealt{Vestergaard06}; \citealt{Shen13}). The SE and RM measurements are largely consistent within their (large) uncertainties for many quasars; however, there is noticeable scatter around a 1-to-1 relation. Our \civ \ lags are consistent with the previously measured \radlum\ relation from which the SE estimators are derived, so we are unsurprised to see so many that are consistent; however, given the uncertainties around \civ\ SE \mbh estimates (see Section~\ref{introduction}), we are also unsurprised to see cases with inconsistencies. A detailed analysis of the reliability of SE mass measurements is beyond the scope of this work, and will be addressed thoroughly in future work dedicated to improving SE mass estimators. 

\section{Summary}
\label{sec:summary} 
With four years of spectroscopic and photometric data from the SDSS-RM program, we searched for time delays between the continuum and the \civ \ emission-line in 349 quasars. Our main results are:
\begin{enumerate} 
\item We measured significant positive lags in \numlags quasars, with an expected false-positive detection rate of 10\%. Lowering the false-positive rate threshold will yield more significant positive lags, but with increased false-positives; including additional years of SDSS-RM monitoring will likely decrease the false-positive rate and lead to a larger set of lags (see Section~\ref{sec:significance}). 
\item We assigned quality ratings to each individual measurement based on visual inspections, leading us to create a ``gold sample" of \numgold of our highest-confidence lag measurements (see Section~\ref{sec:goldsample}). These measurements are consistent with the larger primary sample of \numlags quasars, but are less likely to be false positives and so are the best sources for targeted follow-up of individual quasars. We note again that the criteria used to determine this sample are subjective and thus caution against statistical interpretations using the gold sample. 
\item We place our measurements on the \civ \ \radlum\ relation, which fill in a previously unexplored range of luminosities and increase the number of sources included from 15--18 to $\sim$70 (Section~\ref{sec:rl}). We fit a new relation to our data while including the entire set of \civ \ RM results from the literature, and find relation consistent with previous studies. We separately fit only the gold sample together with previous measurements and measure a consistent relation. We caution that selection effects must be addressed before this relation can be widely used for other applications (such as designing SE mass recipes).
\item We use our time-lag measurements to obtain \mbh measurements for our full sample of lags (see Section~\ref{sec:mbh}). These \mbh values are at the high end of the distribution of RM mass measurements. 
\item We have increased the sample of quasars with \civ \ RM lag measurements from $\sim$18 to $\sim$70, adding quasars at redshifts ranging from 1.35 to 2.8. This is a significant increase in both sample size and redshift range spanned by the RM sample, demonstrating the utility of multi-object RM campaigns in expanding the parameter space covered by RM observations. 
\end{enumerate} 

We have shown here that RM measurements in quasars at higher redshifts and higher luminosities are possible using large survey-based datasets such as ours that span multiple years. Our work makes use of four years of spectroscopic monitoring with SDSS combined with accompanying photometry from the Bok and CFHT telescopes. The SDSS-RM program will continue to observe through 2020 as a part of the SDSS-IV program, and RM monitoring will continue through 2025 as a part of the SDSS-V Black Hole Mapper program (\citealt{Kollmeier17}). The additional years of data will allow us to measure lags in quasars at higher luminosities and explore the SMBH population at unprecedented scales. In addition, we are also adding 4-year PanSTARRS1 early light curves (2010-2014) for SDSS-RM quasars to effectively extend the baseline to measure longer lags (Shen et al., in prep). 

Beyond the SDSS-RM program and the upcoming Black Hole Mapper survey, there are several additional surveys and facilities that are planning or currently executing large RM programs using multi-object spectrographs, such as OzDES (\citealt{King15}), 4MOST (\citealt{Swann19}), and the Maunakea Spectroscopic Explorer (\citealt{McConnachie16}). The SDSS-RM program, and our results here, serve as a proof-of-concept that such programs are not only feasible, but can have a dramatic impact on our knowledge of quasars and SMBHs across the observable universe.

\acknowledgments CJG, WNB, JRT, and DPS acknowledge support from NSF grant AST-1517113. YS acknowledges support from an Alfred P. Sloan Research Fellowship and NSF grant AST-1715579. KH acknowledges support from STFC grant ST/M001296/1. PBH acknowledges support from NSERC grant 2017-05983.

This work is based on observations obtained with MegaPrime/MegaCam, a joint project of CFHT and CEA/DAPNIA, at the Canada-France-Hawaii Telescope (CFHT) which is operated by the National Research Council (NRC) of Canada, the Institut National des Sciences de l'Univers of the Centre National de la Recherche Scientifique of France, and the University of Hawaii.  The authors recognize the cultural importance of the summit of Maunakea to a broad cross section of the Native Hawaiian community. The astronomical community is most fortunate to have the opportunity to conduct observations from this mountain. 

Funding for the Sloan Digital Sky Survey IV has been provided by the Alfred P. Sloan Foundation, the U.S. Department of Energy Office of Science, and the Participating Institutions. SDSS-IV acknowledges
support and resources from the Center for High-Performance Computing at
the University of Utah. The SDSS web site is www.sdss.org.

SDSS-IV is managed by the Astrophysical Research Consortium for the 
Participating Institutions of the SDSS Collaboration including the 
Brazilian Participation Group, the Carnegie Institution for Science, 
Carnegie Mellon University, the Chilean Participation Group, the French Participation Group, Harvard-Smithsonian Center for Astrophysics, 
Instituto de Astrof\'isica de Canarias, The Johns Hopkins University, 
Kavli Institute for the Physics and Mathematics of the Universe (IPMU) / 
University of Tokyo, the Korean Participation Group, Lawrence Berkeley National Laboratory, 
Leibniz Institut f\"ur Astrophysik Potsdam (AIP),  
Max-Planck-Institut f\"ur Astronomie (MPIA Heidelberg), 
Max-Planck-Institut f\"ur Astrophysik (MPA Garching), 
Max-Planck-Institut f\"ur Extraterrestrische Physik (MPE), 
National Astronomical Observatories of China, New Mexico State University, 
New York University, University of Notre Dame, 
Observat\'ario Nacional / MCTI, The Ohio State University, 
Pennsylvania State University, Shanghai Astronomical Observatory, 
United Kingdom Participation Group,
Universidad Nacional Aut\'onoma de M\'exico, University of Arizona, 
University of Colorado Boulder, University of Oxford, University of Portsmouth, 
University of Utah, University of Virginia, University of Washington, University of Wisconsin, 
Vanderbilt University, and Yale University.

We thank the Bok and CFHT Canadian, Chinese, and French TACs for their support. This research uses data obtained through the Telescope Access Program (TAP), which is funded by the National Astronomical Observatories, Chinese Academy of Sciences, and the Special Fund for Astronomy from the Ministry of Finance in China.

\begin{figure*}
\begin{center} 
\includegraphics[scale = 0.335, trim = 0 0 0 0, clip]{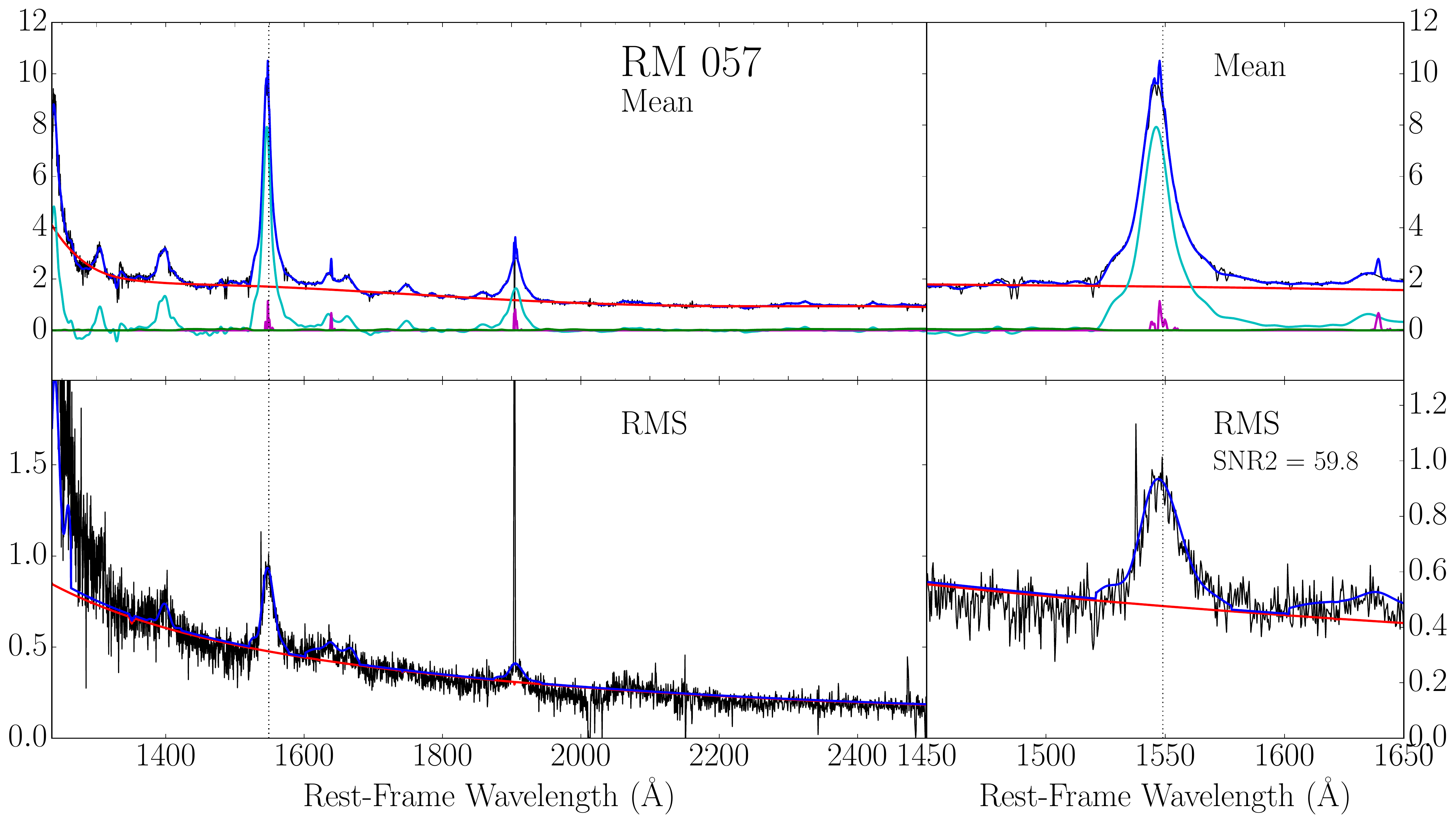}
\caption{Mean and RMS spectra for RM\,057 (SDSS\,J141721.81+530454.3). The top panels show the mean spectrum (black), the continuum fit to the mean (red), the full model fit to the \civ \ emission line (blue), the BLR model (cyan), the Fe\,{\sc ii} model (green), and the narrow-line region model (magenta). The bottom panels show the RMS spectra (black), the RMS model (blue), and the continuum fit to the RMS spectrum (red). Flux densities are in units of 10$^{-17}$ erg~s$^{-1}$~cm$^{-2}$~\AA$^{-1}$. The left panels show a large portion of the observed spectrum, and the right panels show only the \civ \ emission-line region. Vertical dotted black lines indicate the rest-frame wavelength of the \civ \ emission line. Quasars for which significant BALs are present in the fits are denoted with red ``BAL" text in the bottom-right panel. Plots for all \numlags of our quasars with \civ \ lag detections are provided in the online version of the article. } 
\label{fig:meanrms} 
\end{center} 
\end{figure*}

\clearpage
\appendix 
We here present the mean and RMS spectra for our sample of significantly-detected lags (Figure~\ref{fig:meanrms}). In addition, we provide all of the measured quantities used as lag significance criteria for our entire quasar sample (Table~\ref{Table:appendixtbl}).  

\clearpage
\startlongtable
\begin{deluxetable*}{lccccclcr} 
\tablewidth{0pt} 
\renewcommand{\arraystretch}{0.9} 
\tablecaption{Quasar Sample Information} 
\tablehead{ 
\colhead{} & 
\colhead{} & 
\colhead{RA\tablenotemark{a}} & 
\colhead{DEC\tablenotemark{a}} & 
\colhead{} & 
\colhead{} & 
\colhead{} & 
\colhead{} & 
\colhead{} \\ 
\colhead{} & 
\colhead{SDSS} & 
\colhead{(deg)} & 
\colhead{(deg)} & 
\colhead{} & 
\colhead{} & 
\colhead{log $\lambda L_{\lambda 1350}$\tablenotemark{b}} & 
\colhead{log $M_{\rm BH, SE}$\tablenotemark{b,c}} & 
\colhead{} \\ 
\colhead{RMID} & 
\colhead{Identifier} & 
\colhead{(J2000)} & 
\colhead{(J2000)} & 
\colhead{$z$\tablenotemark{b}} & 
\colhead{$i$ mag\tablenotemark{b}} & 
\colhead{(erg s$^{-1}$)} & 
\colhead{(\Msun)} & 
\colhead{SNR2\tablenotemark{d}} 
}  
\startdata 
000 &  J141437.04+530422.7 &  213.6543  &    53.0730 &  1.464 &   20.837    &  44.847$\pm$0.004*  & \nodata &  20.9 \\  
004 &  J141508.57+530019.7 &  213.7857  &    53.0055 &  2.767 &   21.254    &  45.377$\pm$0.003  &  8.47$\pm$0.02 &  20.3 \\  
006 &  J141401.85+530058.5 &  213.5077  &    53.0163 &  1.517 &   21.134    &  44.996$\pm$0.002*  & \nodata &  29.8 \\  
011 &  J141534.20+525743.2 &  213.8925  &    52.9620 &  2.053 &   20.174    &  45.649$\pm$0.001  &  9.09$\pm$0.01 &  42.4 \\  
\smallskip 
012 &  J141355.72+531202.3 &  213.4822  &    53.2006 &  1.585 &   21.499    &  44.740$\pm$0.004*  & \nodata &  30.7 \\  
013 &  J141502.82+525401.2 &  213.7618  &    52.9003 &  1.850 &   21.201    &  44.915$\pm$0.005  &  8.15$\pm$0.02 &  20.6 \\  
019 &  J141529.69+525205.4 &  213.8737  &    52.8682 &  1.918 &   20.117    &  45.422$\pm$0.001  &  8.68$\pm$0.03 &  26.4 \\  
024 &  J141526.06+531941.7 &  213.8586  &    53.3283 &  1.552 &   21.483    &  44.903$\pm$0.002*  & \nodata &  22.7 \\
\enddata 
\label{Table:sample} 
\tablenotetext{a}{These measurements were made as a part of the SDSS Data Release 10 (\citealt{Ahn14}).} 
\tablenotetext{b}{These measurements were retrieved from \cite{Shen18}. $i$-magnitudes listed are PSF magnitudes, and have not been corrected for Galactic extinction. Luminosity measurements with asterisks (*) indicate measurements where $L_{1350}$ was not available; in these cases we converted $L_{1700}$ to $L_{1350}$ using measurements from \cite{Richards06}. } 
\tablenotetext{c}{Black hole mass uncertainties listed here include measurement uncertainties only; the estimated systematic uncertainties beyond those listed is ~0.4 dex.} 
\tablenotetext{d}{SNR2 measurements from PrepSpec (see Section~\ref{sec:sample}).} 
\end{deluxetable*}  

\clearpage
\begin{deluxetable}{lccrr} 
\tablewidth{0pt}
\renewcommand{\arraystretch}{0.8}

\tablecaption{RM\,000 Light Curve} 
\tablehead{ 
\colhead{MJD} & 
\colhead{} & 
\colhead{} & 
\colhead{} & 
\colhead{} \\ 
\colhead{(-50000)} & 
\colhead{Band\tablenotemark{a}} & 
\colhead{Telescope\tablenotemark{b}} & 
\colhead{Flux\tablenotemark{c}} & 
\colhead{Error\tablenotemark{c}}  
}  
\startdata 
      6660.2090 & $g$ & S &       0.99  &       0.06  \\ 
      6664.5132 & $g$ & S &       1.11  &       0.07  \\ 
      6669.5005 & $g$ & S &       1.21  &       0.08  \\ 
      6671.4697 & $g$ & B &       0.93  &       0.18  \\ 
      6671.4717 & $g$ & B &       0.87  &       0.17  \\ 
      6675.4595 & $g$ & B &       1.39  &       0.21  \\ 
      6675.4619 & $g$ & B &       1.46  &       0.20  \\ 
      6675.5303 & $g$ & B &       1.10  &       0.12  \\
\enddata
\tablenotetext{*}{Light curves for all 349 quasars can be found online. A portion is shown here for guidance in formatting.} 
\tablenotetext{a}{CIV = C\,{\sc iv} emission line, $g$ = $g$-band, and $i$ = $i$-band. } 
\tablenotetext{b}{C = CFHT, B = Bok, S = SDSS. } 
\tablenotetext{c}{Continuum Flux densities and uncertainties are in units of 10$^{-17}$ erg s$^{-1}$ cm$^{-2}$ \AA$^{-1}$. Integrated emission-line fluxes are in units of 10$^{-17}$ erg s$^{-1}$ cm$^{-2}$.  } 
 
\label{Table:lcs} 
\end{deluxetable}  
\clearpage

\startlongtable
\begin{deluxetable*}{lc|rrrc} 
\tablewidth{0pt} 
\tablecaption{SDSS-RM Observed-Frame Lag Detections} 
\renewcommand{\arraystretch}{0.9} 
\tablehead{ 
\colhead{} & 
\colhead{} & 
\colhead{$\tau_{\rm JAV}$} & 
\colhead{$\tau_{\rm CCF}$} & 
\colhead{$\tau_{\rm CREAM}$} & 
\colhead{Quality\tablenotemark{a}} \\ 
\colhead{RMID} & 
\colhead{$z$} & 
\colhead{(days)} & 
\colhead{(days)} & 
\colhead{(days)} & 
\colhead{Rating}  
}  
\startdata 
000 & 1.463 & {\color{black} $322.8^{+105.6}_{-90.1}$} & {\color{black} $463.9^{+33.7}_{-163.4}$} & {\color{black} $-675.2^{+48.4}_{-22.6}$} & 2  \\ 
032 & 1.720 & {\color{black} $ 62.0^{+ 9.5}_{- 9.8}$} & {\color{black} $ 57.5^{+61.7}_{-34.8}$} & {\color{black} $ 67.4^{+ 2.1}_{-22.8}$} & 5  \\ 
036 & 2.213 & {\color{black} $605.2^{+50.1}_{-93.1}$} & {\color{black} $416.4^{+104.2}_{-162.5}$} & {\color{black} $601.1^{+30.8}_{-31.9}$} & 1  \\ 
052 & 2.311 & {\color{black} $100.8^{+70.3}_{-20.6}$} & {\color{black} $ 87.0^{+40.4}_{-35.1}$} & {\color{black} $100.3^{+21.7}_{-14.4}$} & 4B  \\ 
057 & 1.930 & {\color{black} $610.4^{+31.2}_{-16.5}$} & {\color{black} $137.6^{+150.0}_{-14.8}$} & {\color{black} $187.1^{+16.8}_{-19.2}$} & 1  \\ 
\smallskip 
058 & 2.299 & {\color{black} $614.0^{+19.5}_{-24.4}$} & {\color{black} $ 84.8^{+62.5}_{-41.8}$} & {\color{black} $177.4^{+43.9}_{-53.0}$} & 1  \\ 
064 & 2.227 & {\color{black} $ 80.2^{+37.4}_{-11.9}$} & {\color{black} $ 80.1^{+58.0}_{-46.9}$} & {\color{black} $ 83.0^{+48.5}_{-10.2}$} & 4B  \\ 
106 & 1.423 & {\color{black} $223.2^{+41.1}_{-31.6}$} & {\color{black} $-683.1^{+38.9}_{-31.7}$} & {\color{black} $-702.9^{+10.7}_{- 5.4}$} & 1B  \\
\enddata 
\tablenotetext{a}{Lag quality rating (see Section \ref{sec:goldsample}). Quasars with significant BAL presence (see Section~\ref{sec:BALs}) are identified with a ``B" following their numerical rating. } 
\label{Table:lags} 
\end{deluxetable*}  

\clearpage
 
\startlongtable
\begin{deluxetable*}{lccrrrrrr} 
\tablewidth{0pt} 
\renewcommand{\arraystretch}{0.9} 
\tablecaption{Line Width, Virial Product, and $M_{\rm BH}$ Measurements} 
\tablehead{ 
\colhead{ } & 
\colhead{ } & 
\colhead{$\tau_{\rm final}$\tablenotemark{b}} & 
\colhead{$\sigma_{\rm line, mean}$} & 
\colhead{$\sigma_{\rm line, rms}$} & 
\colhead{FWHM$_{\rm mean}$} & 
\colhead{FWHM$_{\rm rms}$} & 
\colhead{VP} & 
\colhead{$M_{\rm BH}$\tablenotemark{c}} \\ 
\colhead{RMID\tablenotemark{a}} & 
\colhead{$z$} & 
\colhead{(days)} & 
\colhead{(km s$^{-1}$)} & 
\colhead{(km s$^{-1}$)} & 
\colhead{(km s$^{-1}$)} & 
\colhead{(km s$^{-1}$)} & 
\colhead{(10$^7$ \Msun)} & 
\colhead{(10$^7$ \Msun)}  
}  
\startdata 
000  &  1.463  &  $131.1^{+ 42.9}_{- 36.6}$ & $ 1807 \pm   106$ & $ 2144 \pm    46$ & $ 3509 \pm    74$ & $ 4380 \pm    87$ & $ 11.8^{+  5.8}_{-  5.4}$ &  $ 52.6^{+ 25.9}_{- 24.3}$  \\ 
032  &  1.720  &  $ 22.8^{+  3.5}_{-  3.6}$ & $ 1805 \pm    15$ & $ 2017 \pm    10$ & $ 2768 \pm    22$ & $ 5010 \pm    20$ & $  1.8^{+  0.7}_{-  0.7}$ &  $  8.1^{+  3.2}_{-  3.2}$  \\ 
036  &  2.213  &  $188.4^{+ 15.6}_{- 29.0}$ & $ 2905 \pm    19$ & $ 3900 \pm    34$ & $ 4906 \pm    18$ & $ 7975 \pm   129$ & $ 55.9^{+ 21.1}_{- 22.3}$ &  $249.9^{+ 94.4}_{- 99.8}$  \\ 
052*  &  2.311  &  $ 30.4^{+ 21.2}_{-  6.2}$ & $ 1710 \pm    11$ & $ 1177 \pm    43$ & $ 2164 \pm     7$ & $ 2272 \pm    36$ & $  0.8^{+  0.6}_{-  0.3}$ &  $  3.7^{+  2.9}_{-  1.5}$  \\ 
\smallskip 
057  &  1.930  &  $208.3^{+ 10.6}_{-  5.6}$ & $ 1592 \pm     7$ & $ 1682 \pm    12$ & $ 2652 \pm     8$ & $ 3944 \pm    25$ & $ 11.5^{+  4.3}_{-  4.2}$ &  $ 51.4^{+ 19.1}_{- 19.0}$  \\ 
058  &  2.299  &  $186.1^{+  5.9}_{-  7.4}$ & $ 2695 \pm    24$ & $ 3412 \pm    30$ & $ 3564 \pm    95$ & $ 7512 \pm   121$ & $ 42.3^{+ 15.6}_{- 15.7}$ &  $189.0^{+ 69.9}_{- 70.0}$  \\ 
064*  &  2.227  &  $ 24.9^{+ 11.6}_{-  3.7}$ & $ 1964 \pm    10$ & $ 2431 \pm    17$ & $ 2686 \pm   180$ & $ 3789 \pm    65$ & $  2.9^{+  1.7}_{-  1.1}$ &  $ 12.8^{+  7.6}_{-  5.1}$  \\ 
106*  &  1.423  &  $ 92.1^{+ 17.0}_{- 13.0}$ & $ 5058 \pm   364$ & $ 5414 \pm    36$ & $ 3925 \pm  4017$ & $ 8873 \pm  1322$ & $ 52.7^{+ 21.7}_{- 20.8}$ &  $235.5^{+ 97.0}_{- 93.0}$  \\
\enddata 
\tablenotetext{a}{Quasars with significant BAL inference on the \civ\ emission line (see Section~\ref{sec:BALs}) are flagged with an asterisk. These sources may have incorrect line-width measurements. } 
\tablenotetext{b}{Measurements are in the quasar rest frame.} 
\tablenotetext{c}{Virial products were converted to $M_{\rm BH}$ using $f$ = 4.47, as measured by \cite{Woo15}.} 
\label{Table:mbh} 
\end{deluxetable*}  

\clearpage
\startlongtable 
\begin{deluxetable*}{lrcccc} 
\tablewidth{0pt} 
\renewcommand{\arraystretch}{0.9} 
\tablecaption{Lag Measurements and Significance Parameters for the Entire Sample} 
\tablehead{ 
\colhead{} & 
\colhead{$\tau_{\rm JAV}$} & 
\colhead{Fraction} & 
\colhead{} & 
\colhead{} & 
\colhead{}  \\ 
\colhead{RMID} & 
\colhead{(days)} & 
\colhead{Rejected} & 
\colhead{$r_{\rm max}$} & 
\colhead{SNR$_{\rm con}$} & 
\colhead{SNR$_{\rm line}$} }  
\startdata 
000 & $322.8^{+105.6}_{-90.1}$ & 0.23 &  0.54 & 18.00  &  2.72 \\ 
004 & $194.2^{+34.5}_{-23.5}$ & 0.60 &  0.52 &  8.00  &  0.54 \\ 
006 & $-124.1^{+81.1}_{-139.8}$ & 0.07 &  0.40 & 15.00  &  1.22 \\ 
011 & $245.4^{+89.1}_{-134.7}$ & 0.61 &  0.24 & 14.00  &  0.00 \\ 
012 & $ 13.9^{+19.3}_{-137.5}$ & 0.09 &  0.43 & 11.00  &  2.95 \\ 
\smallskip 
013 & $-430.1^{+55.6}_{-39.6}$ & 0.70 &  0.46 &  7.00  &  0.56 \\ 
019 & $-124.1^{+117.9}_{-149.8}$ & 0.20 &  0.44 &  6.00  &  1.04 \\ 
024 & $524.9^{+112.3}_{-103.4}$ & 0.25 &  0.48 & 11.00  &  2.93 \\
\enddata 
\label{Table:appendixtbl} 
\end{deluxetable*}  

\end{document}